\documentclass[pdflatex,sn-basic,iicol]{sn-jnl}
\usepackage{url}

\RequirePackage{color}
\newcommand\aap{A\&A}                
\newcommand\ao{Appl. Opt.}           
\newcommand\apjl{ApJ}                
\newcommand\apjs{ApJS}               

\newcommand{\fd}{\mbox{$.\!\!^{\mathrm d}$}}

\newcommand{\fp}{\mbox{$.\!\!^{\mathrm p}$}}

\begin{document}

\title[]{Ultraviolet Spectropolarimetry: Conservative and Nonconservative Mass Transfer in OB Interacting Binaries}

\author*[]{\fnm{Geraldine~J.~}\sur{Peters$^{1}$}}\email{gpeters@usc.edu}

\author[]{\fnm{Kenneth~G.~}\sur{Gayley$^{2}$}}

\author[]{\fnm{Richard~}\sur{Ignace$^{3}$}}

\author[]{\fnm{Carol~E.~}\sur{Jones$^{4}$}}

\author[]{\fnm{Ya\"el~}\sur{Naz\'e$^{5}$}}

\author[]{\fnm{Nicole~}\sur{St-Louis$^{6}$}}

\author[]{\fnm{Heloise~}\sur{Stevance$^{7}$}}

\author[]{\fnm{Jorick~S~.}\sur{Vink$^{8}$}}

\author[]{\fnm{Noel~D.~}\sur{Richardson$^{9}$}}

\author[]{\fnm{Jennifer~L.~}\sur{Hoffman$^{10}$}}

\author[]{\fnm{Jamie~R.~}\sur{Lomax$^{11}$}}

\author[]{\fnm{Tomer~}\sur{Shenar$^{12}$}}

\author[]{\fnm{Andrew~G~.}\sur{Fullard$^{13}$}}

\author[]{\fnm{Paul~A.~}\sur{Scowen$^{14}$}}


\abstract{
The current consensus is that at least half of the OB stars are formed in binary or multiple star systems.  The evolution of OB stars is greatly influenced by whether the stars begin as close binaries, and the evolution of the binary systems depend on whether the mass transfer is conservative or nonconservative. FUV/NUV spectropolarimetry is poised to answer the latter question. This paper discusses how the \textit{Polstar} spectropolarimetry mission can characterize the degree of nonconservative mass transfer that occurs at various stages of binary evolution, from the initial mass reversal to the late Algol phase, and quantify its amount.   The proposed instrument combines spectroscopic and polarimetric
capabilities, where the spectroscopy can resolve Doppler shifts in UV resonance lines with 10~km/s
precision, and polarimetry can resolve linear polarization with $\rm 10^{-3}$ precision or better.
The spectroscopy will identify absorption by mass streams and other plasmas seen in projection against the stellar disk
as a function of orbital phase, as well as scattering from extended splash structures, including jets.
The polarimetry tracks the light coming from material not seen against the stellar disk,
allowing the geometry of the scattering to be tracked, resolving ambiguities left by
the spectroscopy and light-curve information.
For example, nonconservative mass streams ejected in the polar direction will produce polarization of the opposite
sign from conservative transfer accreting in the orbital plane.
Time domain coverage over a range of phases of the binary orbit are well supported by the \textit{Polstar} observing strategy.  
Special attention will be given to the epochs of enhanced systemic mass loss that have been identified from \textit{IUE} observations (pre-mass reversal and tangential gas stream impact).
We show how the history of systemic mass and angular momentum loss/gain episodes can be inferred via ensemble evolution through the \textit{r--q} diagram.
Combining the above elements will significantly improve our
understanding of the mass transfer process and the amount of mass that can escape from the system, an important channel for changing the final mass and ultimate supernova of a large number of massive stars found in binaries at close enough separation to undergo interaction.}


\maketitle

\section{Introduction} 
Most massive stars are born in multiple systems. Spectroscopic surveys of the Tarantula Nebula and Galaxy suggest that between half and two thirds of massive stars undergo binary interactions at some point in their lifetimes \citep{sana2012, sana2013}. 
Mass transfer is an important step in the evolution of most massive binary systems: it alters the stellar masses, resulting in stripped stars and rejuvenation, which affects the lifetime and death of the stars. 
As has been called out in the Astro2020 decadal survey, binary interactions are crucial to understand stellar evolution, including the creation of stripped envelope supernovae and kilonovae \citep{Paczynski1971, Langer2012, eldridge2013, yoon2015, tauris2017, Laplace2021}, which are key contributors to the chemical evolution of the ISM and provide an important source of feedback. 
Also, mass lost by a star is not necessarily all accreted by its companion, and escaping mass streams can reduce   
    the final masses of stars by a significant amount that has not been well quantified either observationally or theoretically. In this paper we discuss how the \textit{Polstar} spectropolarimetry mission \citep{ScowenTC} is poised to characterize the degree of nonconservative mass transfer that occurs during close binary evolution, from the initial mass reversal to the late Algol stage, and quantify its amount.

Given the high ionization of massive-star winds, jets, and other circumstellar structures, Thomson scattering (electron scattering) is expected to dominate the induced polarization, producing a useful
sensitivity to the geometry of scattering regions that can be used to resolve ambiguities in the spectroscopic observations.
The classic pioneering model by \citet[][BME hereafter]{1978A&A....68..415B} approximates the time-varying continuum polarization caused by the illumination of stellar winds in a binary system viewed at an arbitrary inclination angle. In this model, the scattering region is described as an optically thin electron gas. The envelope pattern is assumed to co-rotate with the illumination sources, appropriate for steady-state systems in circular orbit. The illuminators consist of one point source at the center of the scattering region and an additional external point source representing the companion. 

\citet{1982MNRAS.198..787B} extended the BME model to consider elliptical orbits, and \citet{1994ApJ...435..372F} further extended the formalism to consider finite illuminators. \citet{1994ApJ...435..372F} showed that occultation is only important in very close binary systems, where separation is less than 10 times the radius of the primary star. However, none of these enhancements to the theory included mass streams involved in
non-conservative mass transfer.   Furthermore, the time and wavelength dependence of the polarization from escaping mass, which could provide unique
constraints and help separate the intrinsic polarization from interstellar polarization (ISP), has not been modeled. 
However, hints of such polarized signals associated with non-conservative mass transfer have been observed in several systems \citep[e.g.$\beta$~Lyr,][]{Hoffman1998,Lomax2012} and extending these observations into the UV, the domain of strongest stellar irradiation
for hot stars, will help quantify the degree of mass loss.  

The optically-thin polarization models are 
crucial since they provide basic insights into information that continuum polarization can provide.
In the case of an axisymetric envelope around
a point source, the observed polarization only depends
on three parameters: the  optical depth $\tau$ of the envelope,
the shape of the envelope (described by a single
parameter $\gamma$ defined below), and the inclination $i$ of the
symmetry axis relative to the observer. If the envelope
is oblate, the observed polarization is parallel to the
symmetry axis. Conversely, the observed polarization is perpendicular to the symmetry axis for a prolate envelope. These results are easily understood:
for a prolate envelope the density is highest along
the poles, and it is these regions --for which the electric vector and
hence polarization is perpendicular to the symmetry axis-- that will contribute most to the observed flux. 
This is one of the ways \textit{Polstar's} polarimetric capability can supplement spectroscopic information, helping to describe
the mass-transfer structure.

In optically thick envelopes, the simple behavior described above is modified.
The orientation is then set not only by where the electric density is highest, it is also set by where the
flux escapes. As a consequence it is possible to
have the polarization switch sign, and, when
the continuous absorption and emissivity is
important, the sign can also be wavelength dependent.
For an axisymmetric system, the polarization is
always perpendicular or parallel to the symmetry axis
(since these are the only unique directions).
Thus, when there is wavelength dependence in the polarization, a signature of switching from
pole-dominated to equator-dominated polarization is a 90 degree rotation in the polarization position angle.
By contrast, a switch from intrinsic-dominated to ISM-dominated polarization can be a rotation of 
the polarization through any position angle.

Line emission offers another probe. For recombination
lines it is typically assumed that the emission
is unpolarized. 
Also, some resonance lines are at least partially depolarizing.
Consequently line photons, which originate at larger distances in the envelope, will
(at least for the optically thin case) show less polarization than the adjacent continuum. 

Our view of the mass transfer process in OB interacting binaries was substantially refined by the pioneering high-resolution FUV spectroscopic observations from the \textit{IUE} and \textit{FUSE} spacecraft (https://archive.stsci.edu/iue/; https://archive.stsci.edu/fuse/). Circumstellar (CS) material was not only in accretion disks, but also in \textit{splash} and other localized outflow regions in the orbital plane, jets that produced mass loss above/below the orbital plane, and high-temperature ($\rm 10^{5}~K$) plasmas. But limited phase coverage within an orbital cycle compromised our ability to map these structures and it was difficult to detect outflows above/below the orbital plane (jets) unless the system displayed a total eclipse. Mass loss from the system during the mass transfer phase was difficult to quantify.

In this paper, we consider how the spectroscopic and polarimetric diagnostics of mass streams and circumstellar disks can reveal the fraction of material that escapes from the system.  
We begin in Section~2 with a discussion of well-known systems, viewed through their location in the \textit{r-q}~diagram.  The latter is of
great importance to our objective of understanding the evolution of mass transfer binaries that are predominantly
B-type systems as they provide a statistically significant sample.
Then in Section~3 we lay out the details of the \textit{Polstar} observational scheme, establishing the required observational requirements
and connecting them to \textit{Polstar's} projected capabilities, to build confidence in the success of an experiment that combines
polarimetry, spectroscopy, and light curve information as a function of orbital phase.
Section~4 continues with a discussion of how polarimetry can be used to infer the mass flow structures in interacting binaries, potentially
containing disks and jets, given their importance for understanding the conservative and
nonconservative elements of the mass transfer budget. 
This section concludes with an application to the well-known B-type binary $\beta $~Lyrae from polarimetric data secured from ground-based instrumentation and FUV spectrophotometry from the spacecraft \textit{WUPPE}.
Section~5 describes the spectroscopic signatures of mass flow, which crucially complements the polarimetric information because of its
quantitative velocity and geometrical constraints on mass streams, splash zones, high-temperature accretion regions, and plasma jets.  
Concluding remarks are given in Section~6.

\section{The Interacting Binary Landscape}\label{rqdiagram}

The importance of binarity on the evolution of OB stars and where the Be stars fit into the picture has been discussed since the mid-1970s \citep{1975BAICz..26...65K, 
    1976IAUS...73..289P, 1976IAUS...70..439P}. The fate of a B star depends on how close it is to its nearest neighbor when it reaches the main sequence.  If the separation of the binary components is small enough, at some point during its post main sequence ($PoMS$) evolution the more massive component will have a radius that is equal to the mean radius of its Roche surface\footnote{These systems are often called {\it close binaries.}}.  Transfer of material from the latter star to its companion will commence during one of the major $PoMS$ expansion phases \citep{1970PASP...82..957P} depending on the initial binary separation: 1) during main sequence expansion (Case A), 2) on the subgiant branch (Case B), or 3) on the asymptotic giant branch (Case C). Although several subgroups of early-type interacting binaries can be associated with the evolutionary groups mentioned above, here we focus on the B-type near contact systems (initial periods up to  $\rm \lesssim 1.5^{d}$ ) and the Algol-type binaries (initial periods from $\rm 1.5^{d} \lesssim P \lesssim 60^{d}$, if the mass loser was originally about 
    $\rm 5 M_{\odot }$ and the mass ratio 0.3).
Typically the contact systems are associated with Case A or slightly beyond and the Algols are usually Case B mass transfer.
 
 \begin{figure*}[ht!]
    \centering
     \includegraphics[width=12.0cm]{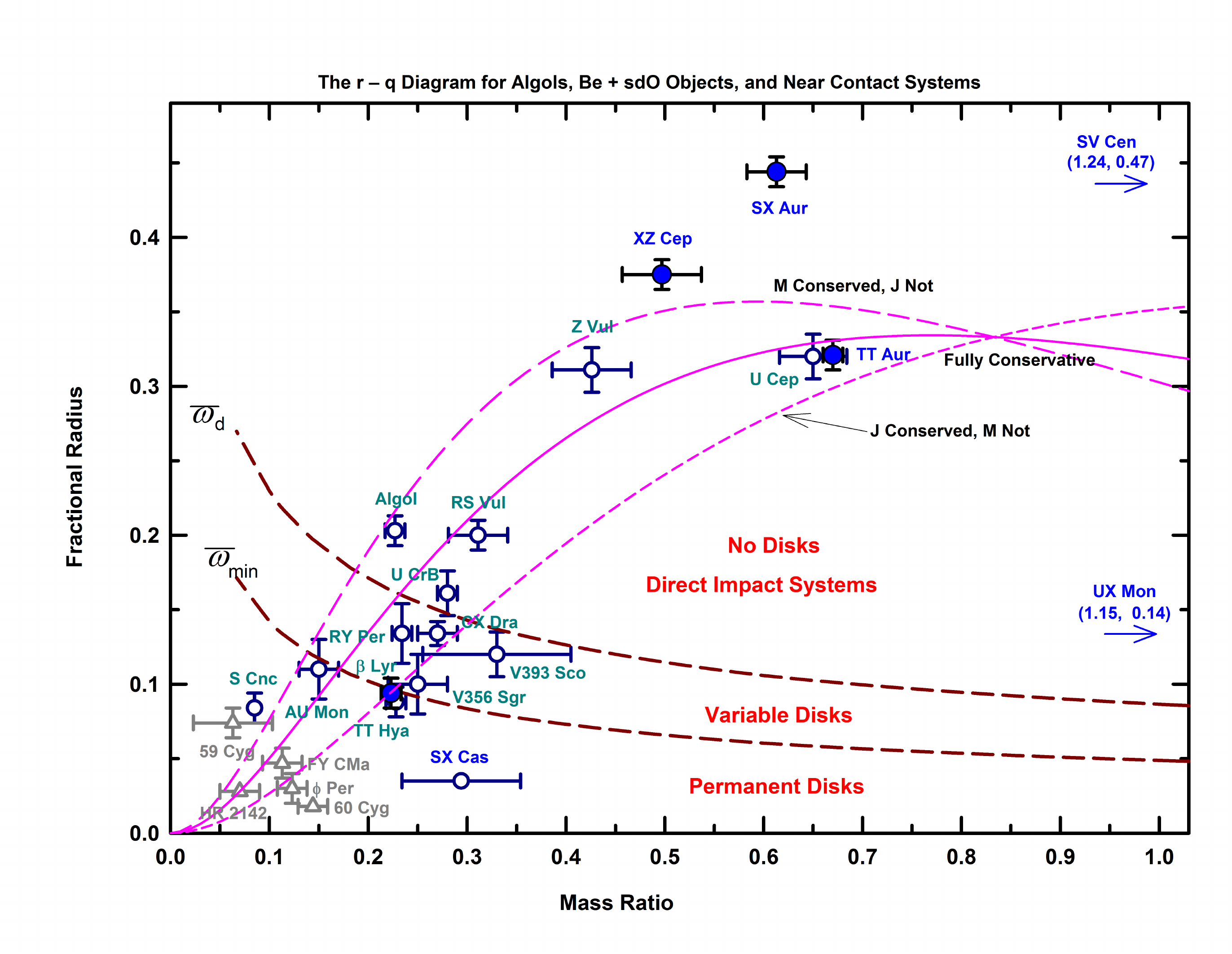}
    \caption{The locations of targets selected for this study (cf. Section 3) in the  \textit{r-q}~diagram. The near contact systems are plotted with \textit{solid blue} circles, while the Algol binaries are indicated by \textit{open} circles and \textit{cyan} labels. The objects plotted with \textit{open gray triangles} are Be +sdO systems. The error bars are from individual papers on each star. The \textit{magenta} long-dashed
    and short-dashed curves indicate, respectively, trajectories of predominantly nonconserved angular momentum and nonconserved mass transfer, as discussed in the text. The evolutionary track for fully conservative mass transfer is shown by the solid \textit{magenta} curve.}  
     \label{fig:rqdiagram}
\end{figure*}

The early calculations on the evolution of close binaries were invariably based upon conservative mass transfer in which all the mass that was lost from the donor star ended up being accreted by the mass-gaining star. The consensus now is that significant mass and angular momentum is indeed lost from the system. The primary objective of this investigation is to identify the stages of mass transfer during which most mass is lost and measure it from phase-dependent FUV spectroscopy and spectropolarimetry.

\citet{1975ApJ...198..383L,1976ApJ...207L..53L} provided us with our first basic understanding on how a localized stellar wind at a binary’s inner Lagrangian point, L1, becomes a defined gas stream that either impacts the mass-gaining star\footnote{In contemporary terminology the star that is losing mass is often called simply the \textit{loser} and the mass-gaining star the \textit{gainer}.} or feeds an accretion disk around it. 
The nature of the CS structures in interacting binaries can be understood with the aid of the \textit{r-q}~diagram in which the fractional
radius of the mass gainer ($\rm R_{p}/a$) is plotted versus the mass 
ratio, $\rm q=M_{loser}/M_{gainer}$, and compared with the theoretical computations of gas stream hydrodynamics of \citet{1975ApJ...198..383L}.  Such a
display is presented in Figure~\ref{fig:rqdiagram}.  The upper \textit{dashed} curve, 
$\rm \varpi_{d}$, delineates the fractional radius of the dense accretion 
disk for systems of different mass ratios.  The lower \textit{dashed} curve, 
$\rm \varpi_{min}$, shows the minimum distance that a stream particle will
achieve relative to the center of the gainer. If the system falls above the 
upper curve (henceforth called \textit{Region 1}), the gas stream will strike the photosphere of the gainer 
and any H$\alpha$ emission seen from the CS material will be transient. These binaries are often called \textit{direct-impact} systems. 
If the system falls below the lower curve (\textit{Region 3}), then one will see 
a prominent accretion disk that emits strongly in H$\alpha$. The amount of 
material in the accretion disk can vary, especially for systems near the 
lower curve.  The most variable accretion disks are found in systems that fall between the two curves (\textit{Region 2}) where the gas stream impact is nearly tangential.  The inner part of a wide gas stream will strike or graze the photosphere of
the primary while the outer edge will feed an accretion disk.   For the
systems plotted in Figure~\ref{fig:rqdiagram} above the $\rm \varpi_{min}$ line, the angle between 
the impacting gas stream and the photosphere of the B star varies from 
$\rm \sim50^{o}$ (e.g. Z~Vul) to tangential (e.g. V393~Sco, V356~Sgr).

After its onset, mass transfer proceeds on a fast timescale until the masses of both stars are the same \citep{1970PASP...82..957P}. If mass
and angular momentum were to be conserved, the orbital period and binary separation would steadily decrease, which would cause a runaway
that would likely violate the conservative assumptions.  Hence we expect substantial mass to be lost from the system at this stage. All or some of the systems may experience Common Envelope Evolution \citep{2013A&ARv..21...59I} during their pre-mass reversal stage, but more studies are needed to understand this phase.  

There are very few recognized examples of pre-mass reversal objects. We consider three such systems here: SX~Aur \citep{1988ApJ...327..265L}, SV~Cen \citep{1982A&A...110..246D,1991ApJ...379..721L}, and UX~Mon \citep{2011A&A...528A.146S}. The radius of the primary component in the SX Aur system is quite close to being in contact with its Roche surface, while its companion is unevolved. This system appears to be on the verge of beginning the mass transfer process. Although the more massive star did not show any obvious evidence of CS material in the \textit{IUE} spectra, an image from the WISE spacecraft \citep{2015A&A...577A..55D} clearly reveals evidence of earlier mass loss.  

SV Cen and UX Mon are definitely in the pre-mass reversal phase. SV Cen is distinguished by displaying the fastest known rate of period decrease, 
\citep[$\rm \dot{P}/P\approx-1.5\cdot 10^{-5}~yr^{-1},$][]{1993IBVS.3868....1D}.
From 1894 to 1993 the period shortened on the average by $\rm 2.2~s~yr^{-1}$,  indicating a mass transfer rate of $\rm 4 \times 10^{-4}~M_{\odot}~yr^{-1} $. 
\citet{1982A&A...110..246D} report systemic mass loss in SV~Cen from the analysis of \textit{IUE} spectra. UX~Mon currently shows a decreasing orbital period of $\rm \dot{P} = -0.260~s~yr^{-1} $ \citep{2011A&A...528A.146S}. The primary may be embedded in an optically-thick accretion disk and the system seems to be related to the W~Serpentis binaries.

All of the Algol binaries shown in Figure~\ref{fig:rqdiagram} were observed with \textit{IUE}, except the O-type system XZ~Cep. Some observational details are highlighted in Section~5. The mass gainers for all of the Algol binaries plotted in Regions 2 and 3 are rotating supersynchronously, as are about half of the systems in Region 1. This is not a surprise because the gas stream is impacting the gainer's photosphere or inner disk with a velocity of about $\rm 400~km~s^{-1}$. It is thought that the systems that are tidally-locked are experiencing a hiatus in the flow of the gas stream. Considering the high velocity of the gas stream, if the impact is tangential one might expect that the material in the flow will head out of the system near phase 0.5 and be lost to the ISM. In fact this is observed: see the discussion of the UV spectropolarimetry of $\beta$ Lyr from \textit{WUPPE} \citep{Hoffman1998,Harmanec1996} discussed in Section 4.

The five longest known Be + sdO (BeS) systems, which are in a post-interaction stage, are seen in the lower left-hand  corner of the \textit{r-q}~diagram in Figure~\ref{fig:rqdiagram}. 
These systems are relevant to the \textit{Polstar} objective \textit{S4} as providing context for the final stages of a
nonconservative transfer process happening earlier, but they are also relevant to another objective, \textit{S3}, which tests the hypothesis that classical Be stars are a post mass transfer stage that eventually produces a BeS object \citep{JonesTC}.
The sdO objects \citep[cf.,][and references therein]{2021AJ....161..248W} are thought to be the stripped down, CNO processed core of a mass loser that transferred significant mass and angular momentum to the Be star. Consider the following scenario. 
If the mass from the loser arrives with too much angular momentum, it must shed
some of its extra angular momentum to allow for more accretion without overloading the gainer. Recall from discussion above that the component separation and orbital period of mass-transferring binary steadily decreases until the mass ratio is 1.0.  After the mass reversal, the opposite happens. 

There is an initial rapid and unstable mass transfer that could have significant mass loss, and then later once stability sets in and both stars thermally equilibrate, we have transverse mass streams and a second phase of mass loss during region 2.  
As the system enters Region~3 the excess angular momentum spins up the gainer (at least its upper envelope and photosphere) and transfers it to a disk structure. The accretion disk material loses energy due to viscosity, so it falls to the photosphere, while the angular momentum is transported outward by viscosity and 
may contribute to an unstable decretion (mass loss) disk. 

This continues as the mass loser shrinks below its Roche surface and exposes its core. Since the Be star in the known bright BeS systems typically has a mass $\rm > 8-9 M_{\odot}$, and the sdO object is $\rm \approx 1.5~M_{\odot} $, if there is substantial systemic mass loss, the original mass loser and current sdO was probably a late O-star.  More information on the sdO objects and their associated Be stars can be found in Jones et al. 2022 (this Topical Collection).

The spinup of the gainer requires only a small amount of angular momentum. The lions share is locked in the binary orbit, and our
current purpose is to understand how much of that angular momentum and total mass is lost from the system during interaction.
For this, we focus on the nonconservative interaction phases alluded to above, and their immediate aftermath.
To focus our attention on the possibilities, there are three theoretical trajectories that curve from the upper right to the lower left in Figure~\ref{fig:rqdiagram}.
The central trajectory is for conservation of both mass and angular momentum, taking as an example a ratio of gainer radius
to orbital separation at mass reversal ($q = 1$) of 1/3.
Note that the largest possible value of this ratio at $q = 1$ is 1/2, because otherwise both stars would need to fill their
Roche lobes at mass reversal, and the assumption is that only the mass loser is undergoing Roche lobe overflow.
Hence the central trajectory is already a fairly aggressive assumption about how large $R/a$ can be without angular momentum
loss from the system.
The \textit{long-dashed} trajectory assumes that no significant mass is lost from the system, but no angular momentum transfer occurs (simulating a limit where the minority of the mass carries off the majority of the angular momentum, such as for a decretion disk). Loss of orbital angular momentum causes the system to spiral inward toward lower separation $a$, raising $r/a$ as shown in Figure~\ref{fig:rqdiagram}.
Hence the systems found at high levels of $r/a$ are likely to require angular momentum loss.

Conversely, the \textit{short-dashed} trajectory below the conservative one assumes the mass is completely lost from the system, but completely conservative angular momentum transfer is occurring (simulating a limit of mass jets being lost that carry little angular momentum).
Loss of mass causes the system to spiral outward toward higher separation $a$, lowering $r/a$. Hence the systems found at low levels of $r/a$ are likely to have experienced significant mass loss without significant
angular momentum loss.

The figure suggests that both types of nonconservative evolution might be occurring in different systems, but this is somewhat
ambiguous because we do not know the history of the systems. We do not know the values of their $r/a$ when $q$ equaled 1.0.  
However, what is clear is that the systems with the highest $r/a$ must have required angular momentum loss, because their
conservative trajectory would track back to $r/a > 1/2$ at $q = 1$, which is not possible given the assumptions
about the mass transfer process as mentioned above.
Given that mass loss that carries a large amount of angular momentum, as well as mass loss that carries only a small amount,
are both potentially playing out in these systems, we wish to watch all these systems carefully, both spectrally and
polarimetrically, at multiple orbital phases,
to look for signatures of either geometry of nonconservative mass transfer in the experiment described next.

\begin{figure*}[ht!]
     \includegraphics[width=16.0cm]{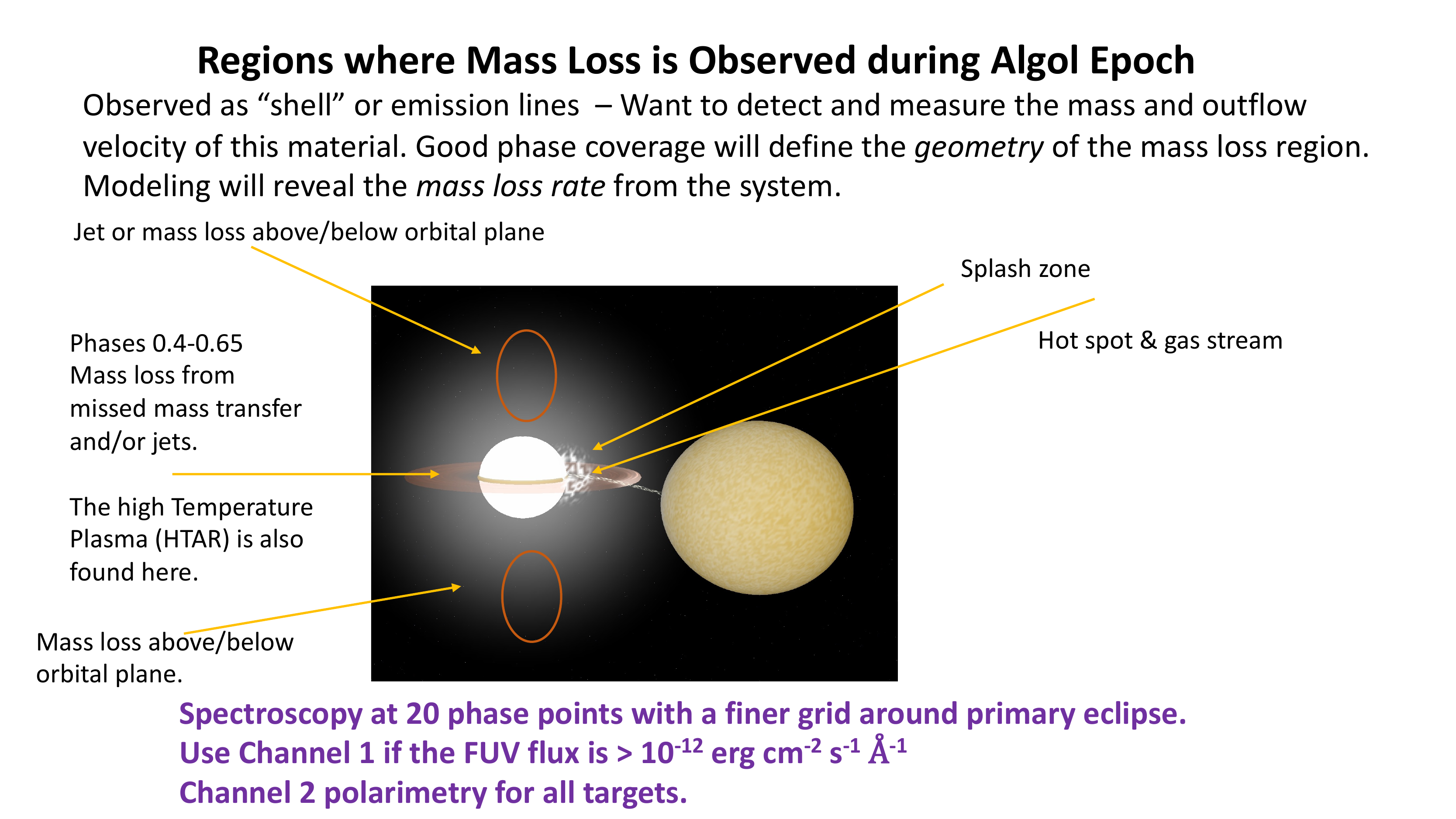}
    \caption{Circumstellar gas/plasma structures observed in Algol-type binaries. All have been inferred from phase-dependent FUV spectral lines and polarimetric signatures. }  
     \label{fig:ryperfig}
\end{figure*}

\section{Proposed Observing Program and Target List}

The \textit{Polstar} observing strategy, instrument requirements, and projected performance is described in this section,
as well as the target list (cf. Table~1) that gives sufficient coverage of key systems to achieve the objective denoted \textit{S4}.
This objective is to look for spectropolarimetric evidence in currently interacting binaries that can test the hypothesis that
mass transfer is significantly nonconservative in its early and rapid phases prior to and shortly following mass reversal. 
The strategy involves 20 visits (10 for the dimmest targets), mostly spread evenly over the phase of the binary orbit but with a few
extra visits over the eclipse phase for those systems that do eclipse.
The plan is to combine spectroscopic searches to resolve absorption lines from mass streams seen in projection against the face of
the primary gainer, with polarimetric data capable of exposing the presence of free-electron scattering well away from the face of the star.
In this way, both the mass transfer rate can be characterized (spectroscopically), as well as the rate of escaping mass (polarimetrically).

Figure~\ref{fig:ryperfig} shows the locations of the most commonly observed CS structures in mass transfer binaries of the Algol type. The disk parameters are usually  determined from modeling hydrogen emission lines (mostly H$\alpha$). But all of the other structures have been identified from modeling FUV absorption spectra from \textit{IUE}, \textit{FUSE}, or \textit{HST}. These structures include the gas stream, a hot spot at the impact site, a \textit{splash} plasma, mass outflow near phase 0.5 (superior conjunction  of the mass-losing star), jets, and a high-temperature plasma in the equatorial region downstream from the impact location.

To carry out the S4 objective, the spectroscopy must be of high enough resolution to pick out mass streams at speeds of a few hundred $\rm km~s^{-1}$,
given that the projection toward our line of sight might be smaller. Many data points are needed across the profiles to study their shape.
Opacity in key resonance lines such as C~IV at 155~nm and N~V at 124~nm is high, as the strength of the resonance compensates for low abundance.
Hence if about 10 data points can be placed across a transient absorption profile that appears cyclically with orbital phase, with
a resolution of about $R \cong 30,000$ one can find the nature of a mass transfer flow from spectroscopy.        
Since the mass transfer rate is high and the infalling velocity gradient weak, 
strong UV resonance lines in the stream will be optically thick.
But since the Roche lobe overflow region is expected to be rather compact, the stream may not cover a large fraction of the face of the 
mass gainer, as Figure~\ref{fig:rqdiagram} shows, so that the gainer may fill a significant fraction of the interbinary separation.
Relatively high SNR ($> 50-100$) is therefore expected to be required to quantify the highly 
unsaturated depth of the absorption by the mass stream in each $\cong 10$ km~s$^{-1}$
resolution element, when it is caught in projection against the stellar disk.
To assist with this desired geometry, high inclination (often eclipsing) systems are chosen in the target list.
In all, the instrument requirements to carry out the spectroscopy of objective \textit{S4} are set at \textit{R~=~30,000} and \textit{S/N~=~50-100}.

The required polarization precision is set at $\rm 1\times10^{-3}$, with the expectation that nonconservative mass transfer
will be indicated by a highly asymmetric geometry of escaping mass.
Systems shown in Figure~\ref{fig:rqdiagram} that exhibit surprisingly wide separations may require an excess of mass loss over angular
momentum loss to achieve such wide separations (as suggested by the short-dashed curve). These systems
could be expected to exhibit mass loss from the system in polar directions, as shown in Figure~\ref{fig:ryperfig}.
Polarization with a position angle aligned with the orbital plane can be anticipated from these systems if the \textit{S4} hypothesis is correct.  

Conversely, systems from Figure~\ref{fig:rqdiagram} that exhibit surprisingly small separations may require high angular momentum loss relative to mass loss, to
cause the orbit to spiral inward suitably to explain their position in the \textit{r-q} diagram.
These latter systems
could be expected to exhibit mass loss from the system in the equatorial plane, seen spectroscopically 
as blueshifted ``shell'' lines at orbital
phases where they cross the stellar disk, and associated with a polarization position angle that is perpendicular to the plane of the orbit.

Since the optical depth of the free-electron scattering responsible for this continuum polarization is significantly less than
the optical depth in strong UV resonance lines, yet the mass flux of significantly nonconservative transfer would rival that of
the mass transfer rate, the free-electron optical depth is expected to be substantial, perhaps comparable to free-electron optical depths ($\sim 0.1$) in the strongest O-star winds.
Such highly aspherical free electron structures would induce polarized scattering, but at a significantly reduced level from their optical depths, so we set the
required polarization precision necessary to detect these escaping mass streams at  $1\times 10^{-3}$. 
This strategy takes advantage of the fact that Ch2 has a roughly factor 3 advantage in effective area over Ch1, such that a factor 2.5 increase in photon count is achieved by only a factor 1.5 increase in total exposure time per visit.  It also allows us to monitor in real time the consistency between the 
channels.


To establish that the projected performance of the \textit{Polstar} experiment can meet these requirements, the target list
in Table~1 shows the required exposure times in Ch1 needed to achieve the listed S/N and the precision in Ch2 for each visit and the number of phase points to be covered. 
The strategy for determining the indicated polarization precision involves combining each Ch1 exposure with a Ch2 exposure of half the Ch1 exposure time.
This takes advantage of the fact that Ch2 has an effective area that is about a factor of three over Ch1,  such that a factor 2.5 increase in photon count is achieved by only a factor 1.5 increase in total exposure time per visit.
This is deemed a useful strategy for helping to build the stringent polarization precision desired, and is included in the
calculation of the indicated polarization precision in the final column.
We see from Table~1 that the instrument requirements are met with margin by these projected performance benchmarks.
The total observing time implicit in the observing strategy is 27 days, an acceptable resource allocation for one
of twelve objectives of \textit{Polstar} in a 3-year mission timeline, while still accounting for a 50\% duty cycle, as well as potential
time set aside for a guest observing program.

\begin{table*}

\caption{Target List }

    \centering
  {\footnotesize 
        \begin{tabular}{c|c|c|c|c|c|c|c|c|c}
       
 \hline
HD & Star & Spectral & System   & Period  & UV Flux in & Exposure  
& Number &  SNR & Precision \\
Number &  Name & Types    &  Type*    & in  &  erg cm$^{-2}$ s$^{-1}$ \AA$^{-1}$ 
&    Time in Ch1           &  of   &   in     &  in Ch2    \\
 & & & & days & (channel) & (s) & phases & Ch1 & (10$^{-3}$)  \\
        \hline
        \hline
5679  &	U Cep       & B8Ve +     & A   & 2.493   & 1.00x10$^{-11}$ (1) & 5000    & 20 & 100  &  \\ 	
  &	           &	    G8IV  	           &     &         & 1.20x10$^{-11}$ (2) &  & & & 0.5\\
10516 & $\phi$~Per  & B1.5e +     & BeS & 126.673 & 1.50x10$^{-09}$ (1) & 600   & 20 & 400 & \\
 &            &          sdO             &     &	   & 1.50x10$^{-09}$ (2) &   & & & 0.3\\
17034 &	RY Per      & B3Ve + & A   & 6.864   & 4.00x10$^{-12}$ (1) &  6000 & 10 & 70   &  \\					
 &	           &	     F7II-III            &     &         & 3.00x10$^{-12}$ (2) &  & &  & 0.7 \\
19356 &	Algol       & B8V +      & A   & 2.867   & 3.00x10$^{-09}$ (1) &  600  & 20 & 600 & \\								
 &	           &	      K2IV            &     &         & 3.00x10$^{-09}$ (2) &   &  & & 0.2 \\
33088 &	TT Aur	    & B2V +      & EC  & 1.333   & 1.00x10$^{-11}$ (1) & 3000   & 20 & 80 &    \\  						         
 &            &          B4IV        &     &         & 5.00x10$^{-12}$ (2) &   &  & & 0.9  \\
33357 & SX Aur      & B1.5V +     & EC  & 1.210   & 2.00x10$^{-11}$ (1) &   2500 & 20 & 100 &   \\		
 &            &            B3V        &     &         & 6.00x10$^{-12}$ (2) &  &   & & 0.9  \\
41335 & HR 2142     & B1.5Ve +    & BeS & 80.913  & 4.00x10$^{-10}$ (1) &  200 & 20 & 120 &    \\		
 &            &          sdO       &     &         & 2.00x10$^{-10}$ (2) & &  & &  0.6  \\
50846 & AU Mon      & B3Ve +    & A   & 11.113  & 1.50x10$^{-11}$ (1) &  3600 & 20 & 100 &    \\	
      &                   &    F8III           &     &         & 4.00x10$^{-12}$ (2) & &  & &  0.8   \\
58978 &	FY CMa	  & B0.5Ve +    & BeS & 37.257  & 6.00x10$^{-10}$ (1) & 600 & 20 & 250 & \\								
      &                  &    sdO           &     &         & 6.00x10$^{-10}$ (2) &  & & & 0.3  \\ 
65607 & UX Mon     & A5Ve +    & A   & 5.904   & 1.00x10$^{-12}$ (1) &  30000 & 10 & 80 &  \\		
      &                    &   G2III            &     &         & 1.00x10$^{-12}$ (2) & &  & & 0.6 \\
74307 & S Cnc      & B9.5V +    & A   & 9.485  & 4.00x10$^{-12}$ (1) &  6000 & 10 & 70 &   \\		
      &              &    G8IV           &     &        & 4.00x10$^{-12}$ (2) & & & & 0.6 \\
97528 & TT Hya      & B9.5Ve +    & A   & 6.953  & 9.00x10$^{-12}$ (1) &   4000 & 10  & 80&  \\		
      &             &   K3IV            &     &        & 4.00x10$^{-12}$ (2) &  &  & & 0.7 \\
102552 & SV Cen   & B1V +   & EC  & 1.658  & 6.00x10$^{-12}$ (1) &  6000 & 10 & 80 &   \\		
      &             &  B6.5III             &     &        & 4.00x10$^{-12}$ (2) &  & & & 0.7 \\
136175 & U CrB      & B6V +  & A   & 3.452  & 2.00x10$^{-11}$ (1) &   2500 & 20  & 100 &  \\		
      &       &      F8III-IV         &     &        & 4.00x10$^{-12}$ (2) & &   & & 0.7 \\
155550 & FV Sco    & B4IV +     & A   & 5.727  & 1.50x10$^{-11}$ (1) &   4000 & 10  & 100 &  \\		
      &            &   F-G:            &     &        & 5.00x10$^{-12}$ (2) &  &   & & 0.6 \\
161741 & V393 Sco  & B3V +        & A   & 7.713  & 1.50x10$^{-11}$ (1) &   4000 & 10 & 100 &   \\		
      &            &  A:             &     &        & 8.00x10$^{-12}$ (2) &  &    & & 0.7 \\
173787 & V356 Sgr  & B3V +      & A   & 8.896  & 4.00x10$^{-12}$ (1) &   6000 & 10 & 70 &   \\		
      &             &  A2II             &     &        & 7.00x10$^{-12}$ (2) & &     & & 0.5 \\
174237 & CX Dra   & B2.5Ve +  & A   & 6.696  & 2.00x10$^{-10}$ (1) &   900 & 20 & 200 & \\		
       &            &     F5III       &     &        & 8.00x10$^{-11}$ (2) & & & & 0.5\\
180939 & RS Vul	   & B5V +     & A   & 4.478  & 3.00x10$^{-11}$ (1) &   2000 & 20 & 100 &  \\		
      &            &   G1III            &     &        & 1.30x10$^{-11}$ (2) & &  & & 0.6  \\
181987 & Z Vul     & B3V +       & A   & 2.455  & 2.00x10$^{-11}$ (1) &   2500 & 20 & 100 & \\		
      &            &    AIII           &     &        & 3.10x10$^{-11}$ (2) &  &    & & 0.4 \\
200120 & 59 Cyg	   & B1Ve +    & BeS   & 28.187  & 1.10x10$^{-09}$ (1) &   600 & 20 & 350 & \\		
      &         &    sdO        &        &        & 1.10x10$^{-09}$  (1)&   &  & & 0.3 \\
200310 & 60 Cyg      & B1Ve +    & BeS   & 146.6  & 5.00x10$^{-10}$ (1) &   600 & 20 & 200 & \\		
      &             &       sdO     &        &        & 1.70x10$^{-10}$ (2)&  & & & 0.5 \\
      & XZ Cep    & O9.5V +    & O   &  5.1 &   &   3600 & 20 & 100 &  \\		
      &           &    B1III           &     &   &        & & & & 1 \\ 
232121 & SX Cas    & B5Ve +    & W   & 36.561 & 4.00x10$^{-13}$ (1) &   30000 & 10 & 50 &  \\		
      &           &    K3III           &     &         & 1.00x10$^{-13}$ (2) & & & & 1 \\ \hline
      
\end{tabular}

\noindent *A=Algol, BeS=Be+sdO, EC = Early-Type Contact,O=O-type system, W=W~Ser. 
}
   
\label{tab:targetlist}

\end{table*}

\section{Polarimetry of Interacting Binaries with Disks}

\subsection{Basics of polarimetric behavior}

There is a suite of modeling tools, such as Monte Carlo radiative transfer \citep{2017ascl.soft11013W,2017IAUS..329..390C,2018MNRAS.477.1365S}, for evaluating polarization from binaries in optically thin or thick regimes.  It is however useful to explore the thin scattering regime, particularly the theory advanced by \cite{1978A&A....68..415B}.  While thin scattering can be limited in providing quantitatively accurate results \citet[e.g.,][]{1996ApJ...461..828W}, it can often provide appropriate order of magnitude results and especially insights into qualitative behavior.  Moreover, owing to the low value of the Thomson scattering cross-section, there are many applications in which thin scattering applies at the densities representative of some circumstellar media.

Here we review the \cite{1978A&A....68..415B} approach for a binary system using somewhat modified notation and presentation that draws partly on \cite{1978A&A....68..415B} and partly on \cite{1982MNRAS.198..787B}.  The \cite{1978A&A....68..415B} approaches makes several simplifying assumptions such as point source illumination, neglect of occultation effects, no reflections off either star, and thin scattering as already noted.  The authors then proceed to define distributions of the scattering envelope as weighted essentially by lower-term spherical harmonics \citep[e.g.,][]{1982MNRAS.200...91S,1983MNRAS.205..153S}.  There are three broad classifications of these moments:  overall asphericity, left-right asymmetry, and front-back asymmetry.  Overall asphericity is in relation to the orbital plane of the binary; left-right is respect to the line-of-centers (LOC) joining the two stellar components; and back-front is with respect to each star.

\begin{figure}[ht]
    \centering
    \includegraphics[width=7cm]{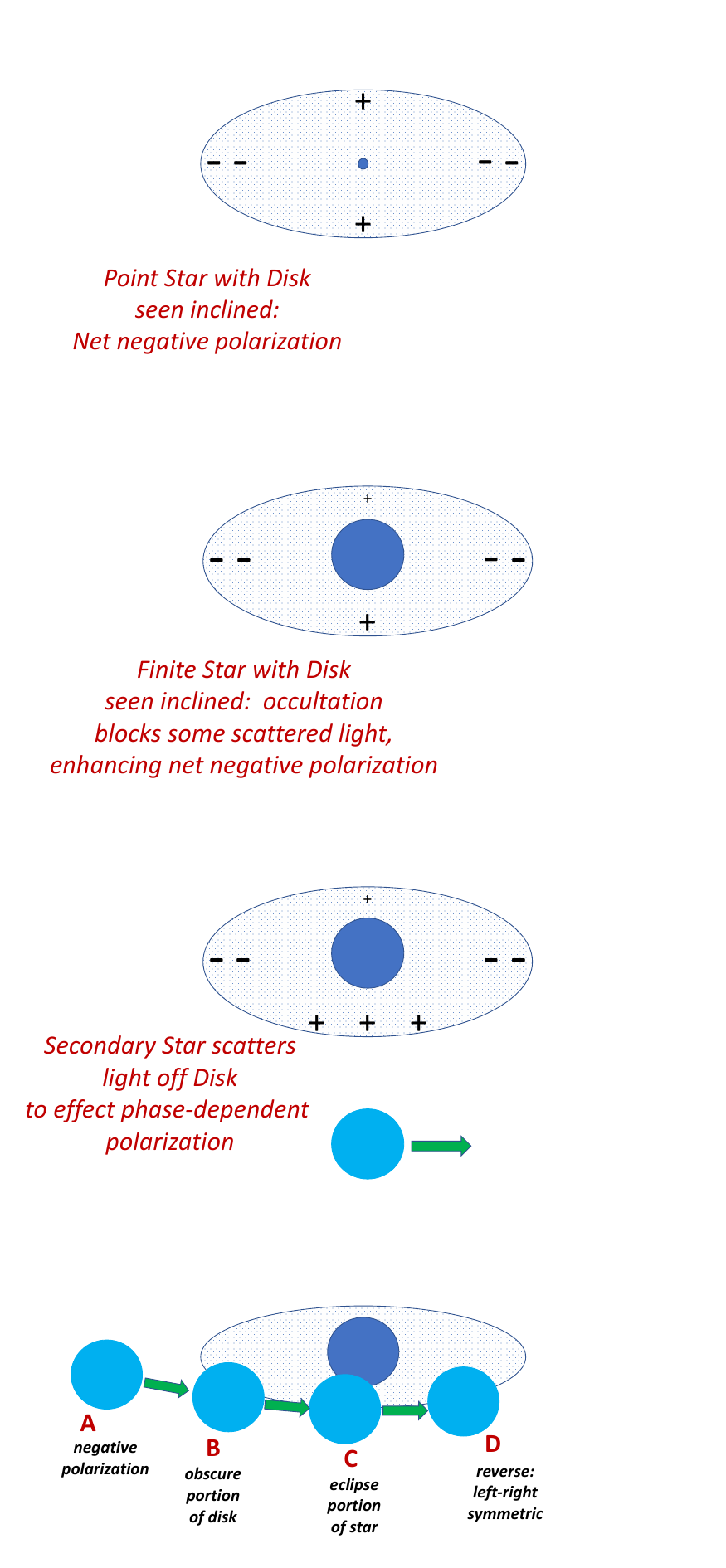}
    \caption{Schematic representation for four scenarios producing constant or time variable linear polarization, as discussed in text.  Top is an axisymmetric disk around a point star.  Next down is the same disk with a finite sized star.  Third is the addition of an orbiting companion star.  Bottom is a higher inclination scenario of relevance for many close binary targets of {\em Polstar} involving eclipse effects.}
    \label{fig:polcartoon}
\end{figure}

As an example, consider the representative schematics of Figure~\ref{fig:polcartoon}.
Top shows a single isotropic point source with a circumstellar disk.  This is aspherical and yields a net polarization when the disk is spatially unresolved.  As long as the disk is steady-state (i.e., no change of the axisymmetric structure or optical depth), the polarization is constant.  The disk is viewed at some inclination, certainly not pole-on but neither edge-on, and the net polarization would be negative by convention:  the net polarization would be oriented with the symmetry axis of the disk.

Allowing for finite stellar size of the star (2nd schematic from top) leads to occultation, which can alter the level of net polarization.  Although occultation is not treated by \cite{1978A&A....68..415B}, the topic has been considered by several authors \citep[e.g.,][]{1991ApJ...379..663F}. The effect of a finite sized star induces no time variable polarization as long as the disk is steady.  Indeed, even if the star itself is variable, the polarization will remain constant being a ratio of polarized flux to total flux.

In the next schematic (3rd from top), a binary companion is added (referring to it as the secondary, whereas the star with disk is the primary), and the polarization becomes time-dependent and cyclical with orbital phase.  Even if the disk is steady-state, both the polarization and position angle (PA) can change as the secondary orbits the primary, owing to the changing orientation of how starlight is scattered into the observer sightline.  Relative to the secondary, the disk represents a front-back asymmetry for the scattering distribution, but not left-right asymmetry.

The bottom schematic in Figure~\ref{fig:polcartoon} is another scenario not treated by 
\cite{1978A&A....68..415B}. In this case the system's inclination is large enough for secondary star to eclipse both the primary and the disk. This is shown for two reasons.  First, several targets for \textit{Polstar} are close eclipsing binaries and eclipses contain a wealth of diagnostic material involving both spectral line and continuum polarization for extracting the geometry of the binary system.  Although \cite{1978A&A....68..415B} only deals with point sources, this last schematic also highlight how in a close binary, one expects accretion and thus an accretion stream.  Indeed one may even expect a hot spot on the disk where mass from the stream enters the disk \citep[e.g., ][]{2012ApJ...750...59L}.  Although the schematic does not include a stream or hot spot, these have two main effects.  

First, one can easily imagine that the stream creates a break in left-right symmetry about the LOC between the stars.  This introduces a lead-lag effect in the polarized light curve relative to eclipsing light curve.  Second, a hot spot will have a spectrum, representing a third source of illumination for scattering by electrons in the ionized envelope.  The approach of \cite{1978A&A....68..415B} allows for any number of illuminating sources to explore such effects, especially wavelength-dependence.

Note that orbital eccentricity can also introduce new effects \citep[e.g.,][]{1982MNRAS.198..787B,2000AJ....120..413M}, but for relatively close and typically interacting binaries, we shall assume circular orbits in presenting heuristic models to illustrate (a) how geometry is extracted from variable continuum polarimetry and (b) chromatic effects in the UV that may improve defining the geometry based on the data.

Monte Carlo radiative transfer approaches are capable of taking into account the many realistic factors that influence variable polarization from interacting binaries:  the effects of finite stellar size for eclipses and occultation, the effects of multiple scattering, the effects multiple components such as a hot spot or accretion stream.
However, for purposes of illustrating the essentials of modeling polarimetric variability from hot star binaries, and in particular the expectation of chromatic effects in the UV band observable with \textit{Polstar}, an example is developed using the approach of \cite{1978A&A....68..415B}.  
Following that paper, adopting some notation from \cite{1982MNRAS.198..787B}, and modifying it slightly, the linear polarization in Stokes-Q and Stokes-U for a binary with a circular orbit is given by

\begin{eqnarray}
q & = & \bar{\tau} -3\tau_0\sin^2 i + G\sin 2i\,\cos(\phi+\phi_G) \nonumber \\
 & & 
    -H (1+\cos^2 i)\,\cos[2(\phi+\phi_H)] \\
u & = & 2 G\sin i\,\sin(\phi+\phi_G) \nonumber \\
& & -2H\cos i\,\sin[2(\phi+\phi_H)] ,
\end{eqnarray}

\noindent where $i$ is the viewing inclination (such that $i=0^\circ$ is
a top-down view of the orbit) and $\phi$ is the azimuth of the LOC in
the orbit plane, relative to the viewer sightline.  The coefficients $G$ and $H$ are given in \cite{1982MNRAS.198..787B}, where $G$ contains terms that are left-right asymmetric, but also top-down asymmetric, while $H$ contains terms that are left-right asymmetric and back-front asymmetric. 
We will assume that the scattering distribution is top-down symmetric, so $G=0$. 
As a result $q$ has a constant term and time or phase dependent term with pattern variation of twice per orbit and potential phase lag signified
by $\phi_H$.  Then $u$ is similar but with no constant term.  The polarimetric variation in a $q-u$ diagram is an elliptical figure based on the variable terms, with two circuits about the ellipse in one orbit, and a constant offset in the direction of $q$.  The eccentricity of the ellipse relates to the viewing inclination, with circular for $i=0^\circ$ and degenerating to a line for $i=90^\circ$ (since $u$ would be zero).

Definitions for the remaining parameters are:

\begin{eqnarray}
H^2 & = & \tau_3^2+\tau_4^2 \\
\tan 2\phi_H & = & \tau_4/\tau_3 \\
\bar{\tau} & = & f_1(\lambda)\bar{\tau}_1 + f_2(\lambda)\bar{\tau}_2 \\
\tau_0 & = & f_1(\lambda)\tau_{01} + f_2(\lambda)\tau_{02} \\
\tau_3 & = & f_1(\lambda)\tau_{31} + f_2(\lambda)\tau_{32} \\
\tau_4 & = & f_1(\lambda)\tau_{41} + f_2(\lambda)\tau_{42} . 
\end{eqnarray}

\noindent where $f_1$ and $f_2$ are the relative monochromatic luminosities of the respective stars (1 for primary and 2 for secondary) as fractions of the total:  $f_1 = L_1/(L_1+L_2)$ and $f_2=L_2/(L_1+L_2)$.  As will be addressed shortly, it is these terms that can allow for chromatic effects despite electron scattering being a gray opacity.  For the various $\tau$ parameters
at left, the additional subscript of 1 or 2 for terms on the right signify moment calculations of the scattering distribution relative to the respective stars.  Note that symbol ``$\tau$'' is chosen because those factors scale as optical depths.  Each one scales as the product $n_0\sigma_T L$, for $n_0$ a characteristic number density of electrons, $\sigma_T$ the Thomson cross-section, and $L$ a characteristic length.  And this holds for the various defined components in the system, such as a disk, an accretion stream, a jet, or a wind. The definitions of the $\tau$ terms can be found in BME in terms of the integral relations (although those authors refer to shape factors $\gamma$, but the integrals are the same).

The offset term involving $\bar{\tau}$ and
$\tau_0$ does not contribute to variable polarization.  It is
an offset, and without additional information, might not be easily
disentangled from the ISP, unless the Serkowski
fit \citep[e.g.,][]{1991ApJ...382L..85T,1992ApJ...385L..53C} is quite well determined.  Note that the offset can show
chromatic effects in the UV, to be addressed later.  It is however
the variable polarization that can be straightforwardly attributed
to the binary as distinct from the interstellar contribution.  Consequently, we ignore the offset in the discussion that follows, although the offset contains additional useful information about the geometry of the stellar system if it can be disentangled from the ISM polarization \citep[e.g.,][]{2001ASPC..233..261N}.

Now to address chromatic effects that have been mentioned several times.
The two fractions $f_1$ and $f_2$ are wavelength dependent.  Since the
focus is on hot massive stars, in the optical and IR band, the continua spectra of both stars will be approximately Rayleigh-Jeans.  As a result, both $f_1$ and $f_2$ become constant, as the $\lambda^{-4}$ factor cancels.  Consequently, the polarization is flat across those spectral bands.  (The polarization is also flat if one star dominates the luminosity at every wavelength of observation.)

However, in the UV, the two stars will generally have different temperatures and therefore different Wien peaks.  The different temperatures and different radii of the stars mean that $f_1$ and $f_2$ also become wavelength-dependent, such that the luminosity weighted terms contribute to $q$ and $u$ to produce polarization with chromatic effects (i.e., not flat polarized spectra).  The chromatic effects emphasize certain moment terms that better define the geometry of the system in relation to model fitting.  In particular, the phase lead-lag ($\phi_H$) is itself chromatic, a strong constraint on geometrical components in the spatially unresolved system.  Note that the moment terms expressed as the $\tau$ parameters are not chromatic but are fixed by the geometry of the system; it is only the weighted sums that are chromatic by virtue of the two different stellar spectral distributions.

\begin{figure}[ht]
    \centering
    \includegraphics[width=7cm]{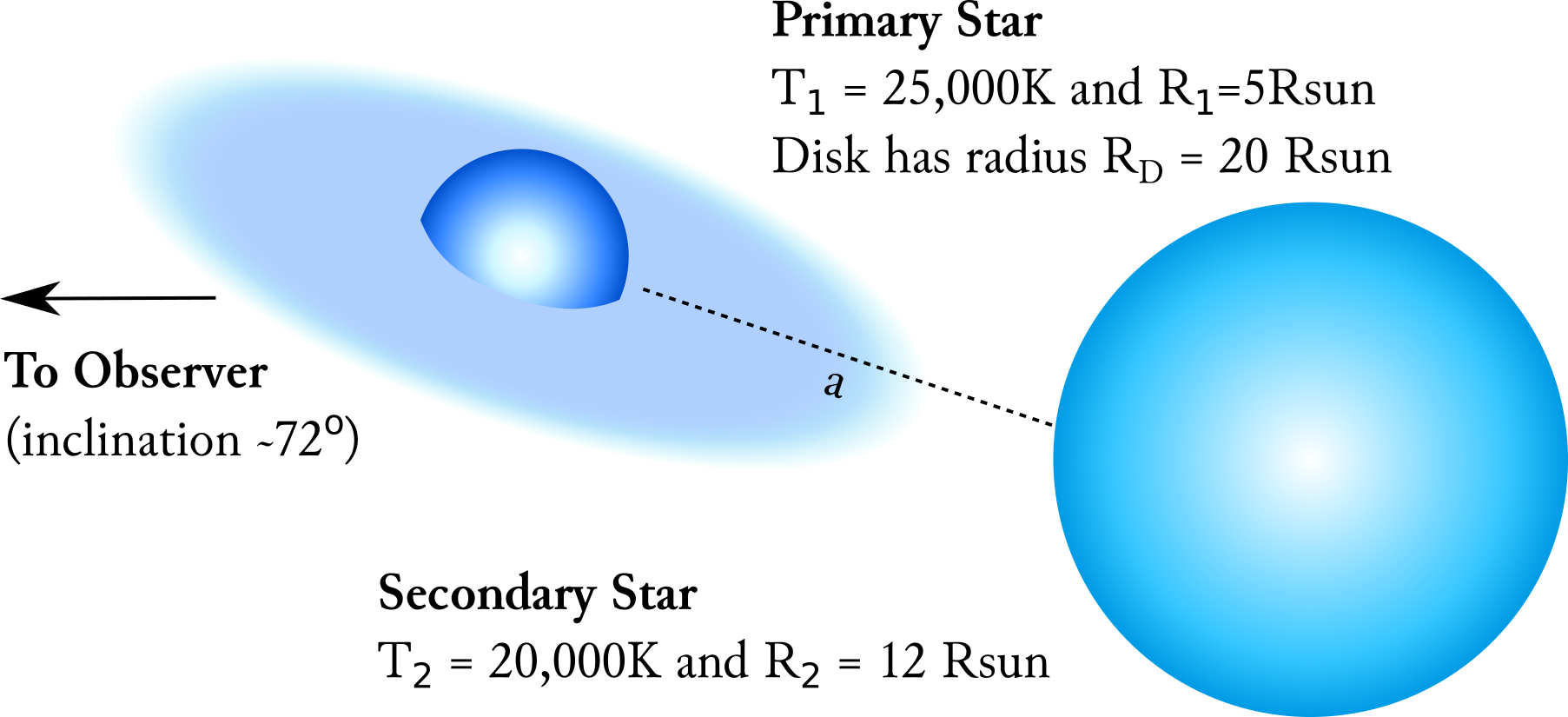}
    \includegraphics[width=7cm]{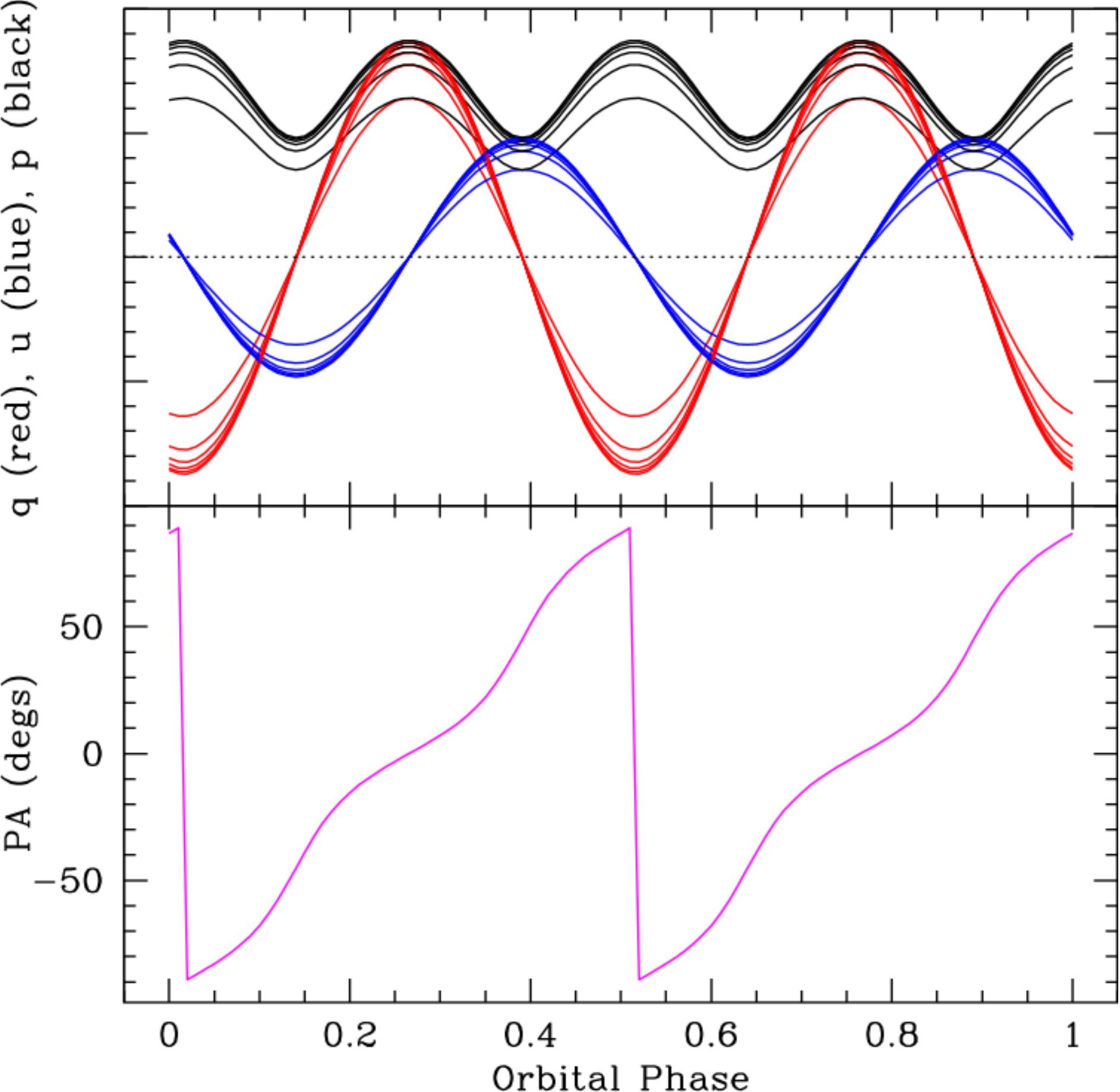}
    \caption{Three panels as illustration of polarimetric variability
    for a binary system.  Top:  A cartoon to depict a primary surrounded by an axisymmetric disk orbited by a secondary star with temperatures and sizes indicated.  Middle:  Variation of $q$, $u$, and $p=\sqrt{q^2+u^2}$ with colors indicated in the label.  Note that the different curves are for different wavelength, from 100nm (smallest $p$) to 400nm (highest $p$).  The variation arises entirely from scattered light by the secondary star, since the disk is axisymmetric about the primary.  Bottom:  The position angle (PA) changes with orbital phase.  Note that in this model there is no phase lag, for any wavelength, because the system is left-right symmetric; the presence of an accretion stream could change that.}
    \label{fig:polmodel}
\end{figure}

For a simple illustration, Figure~\ref{fig:polmodel} shows results using the preceding expressions for a binary involving a primary star surrounded by an axisymmetric disk; see schematic to scale at top.  The primary is hotter and smaller; the secondary is cooler and larger, with parameter values specified in the schematic.  The inclination is about $72^\circ$ ($\cos i = 0.3$).  Polarimetric variability is displayed in the middle panel with colors as specified in the label for $q$, $u$, and $p=\sqrt{q^2+u^2}$.  Finally at bottom is shown the position angle (PA) variation.  Both middle and bottom are plotted with orbital phase, ranging from 0 to 1.  At phase zero, the secondary would be forefront if seen edge-on.

Several key points are noted.  First the various curves in the middle panel are for different wavelengths of 100~nm to 400~nm in equal steps.  For sake of illustration, the stellar spectra of the two stars are taken as Planckian at the indicated temperatures.  For the selected parameters, the lowest polarization is for 100~nm, and highest for 400~nm.  Note how the curves get closer together with increasing wavelength as the individual stars begin to approach their respective Rayleigh-Jeans spectral distributions, just as expected.  

Also, the middle panel has no scale provided.  The polarization from thin scattering scales with the optical depth of the medium.  We can expect amplitude levels for $p$ at $10^{-4}$ to $10^{-3}$ fractional polarizations.  But $p$ is positive definite, whereas $q$ and $u$ are signed.  For edge-on systems we expect $u=0$, and $q$ varies between $\pm p$, implying a factor of 2 gain in relative change of polarization, a gain for detection and monitoring purposes.

For the bottom panel, the PA indeed shows cyclic variations at twice the rate of the orbital period.  This occurs because the geometry is left-right symmetric.  However, whereas the polarization curves in the middle panel show chromatic effects, such effects are absent from the PA variation.  This is because there is no left-right asymmetry in the geometry -- no accretion stream is included in the model.  The accretion stream could introduce a phase lag that is wavelength-dependent.  What would that look like?  At each wavelength the polarization would trace out an ellipse in the $q-u$ diagram.  Each ellipse would be of different sizes for different wavelengths, although the eccentricity would remain constant (since the viewing inclination is constant). The effect of different phase would produce relative rotation shifts between the ellipses.  

It is important that the approach of BME can be expanded further to include orbital eccentricity \citep[e.g., colliding wind systems,][]{StLouisTC,2022ApJ...933....5I} and even a hot spot on the disk.  The latter would involve inserting another source of illumination, thus expanding the diversity of chromatic responses.  The hot spot will not lie on the LOC of the binary stars, and so would introduce yet an additional phase lag.  

The use of BME shows that the UV spectral band is ideal for extracting important information about the geometry of the system, such as symmetry of the disk, or presence of an accretion stream.  When used in conjunction with line profile variability, plus the system light curve (most of our targets are eclipsing variables), the scope of the dataset is rich indeed for extracting densities and kinematics of flows, particularly accretion rates, in the system, and additional component representing non-conservative mass-transfer.  While BME neglects radiative transfer effects (such as multiple scattering), our team possesses the expertise to model such effects, guided by BME for qualitative behavior and as benchmarks for the detailed codes in the optically thin limit. For example, \citet{Hoffman2003} applied MCRT methods to explore the effects of multiple scattering in binary systems that include both interior and exterior disk illumination.

\begin{figure}[ht!]
    \centering
    \includegraphics[width=7cm]{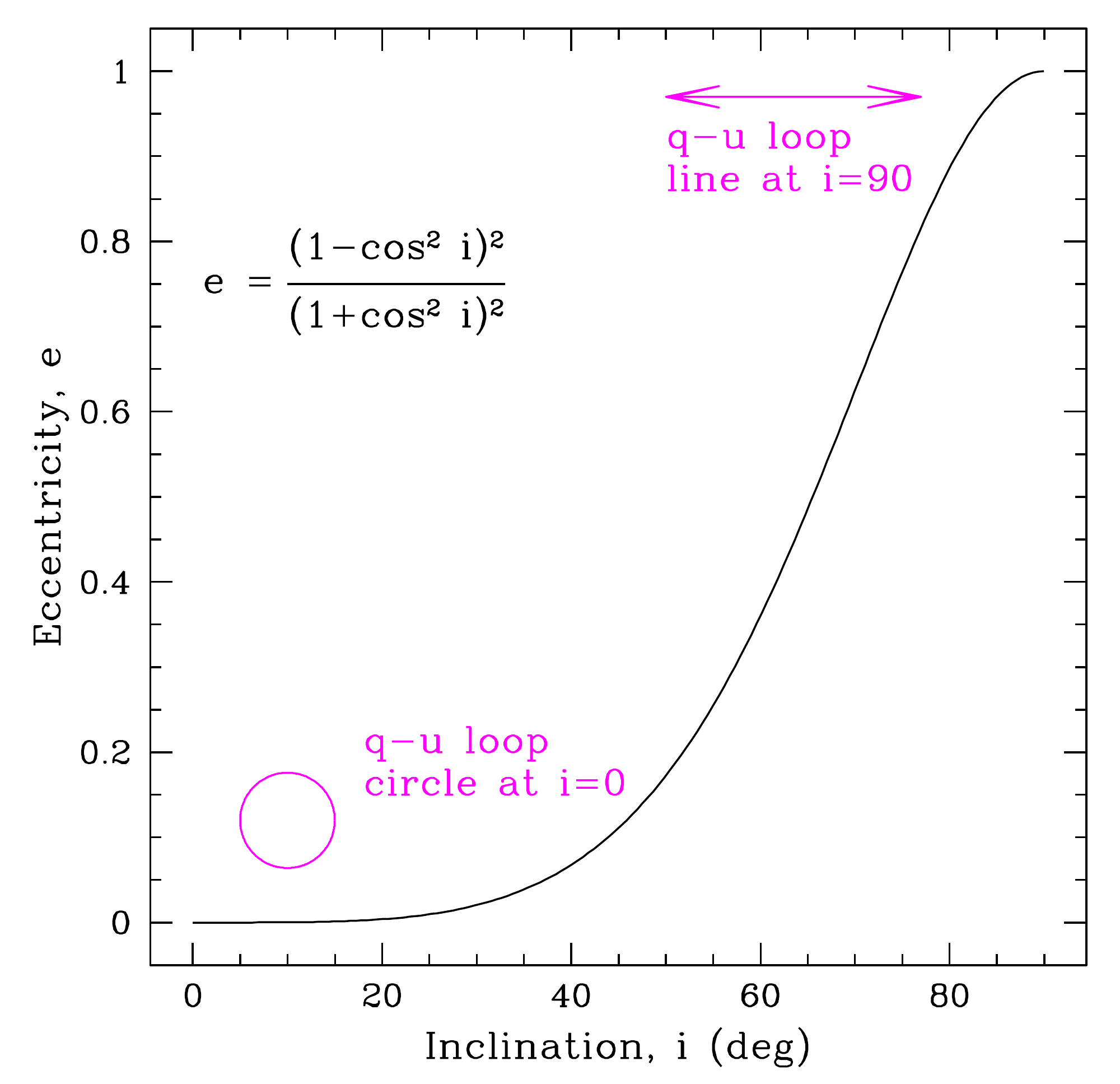}
    \caption{Eccentricity, $e$, for the elliptical variations in a $q-u$ diagram for a binary in a circular orbit plotted against viewing inclination, $i$.}
    \label{fig:ecc}
\end{figure}

While many of the targets are eclipsing binaries, some are not, and polarimetry can provide a measure (or an additional constraint) to the viewing inclination of the binary orbit.  This is important since inclination can often be the greatest ambiguity in the determination of the individual stellar masses.  For elliptical variations, the major and minor axes are set by the extrema $\Delta q$ and $\Delta u$, with $e = 1-(\Delta u^2/\Delta q^2)$.  In relation to the viewing inclination:

\begin{equation}
    e = \left(\frac{1-\cos^2i}{1+\cos^2 i}\right)^2.
\end{equation}

\noindent This relationship is plotted in Figure~\ref{fig:ecc}.  Note that the error on the measured eccentricity will depend on $\Delta q / \sigma_q$ and $\Delta u / \sigma_u$, for $\sigma_q$ and $\sigma_u$ the polarization precision.  At high inclination near edge-on, the effort is aided by the fact that the system will be eclipsing.  At low inclination, the challenge is the curve is relatively flat below $i \approx 30^\circ$.  For fixed measurement uncertainty, the polarimetric variations must be relatively large to distinguish between a $q-u$ loop that is strictly circular (corresponding to pole-on) and one that is only mildly eccentric.  Between about $30^\circ$ to $80^\circ$, the relation is more linear.

\subsection{Observable signatures of mass loss}

\begin{figure*}[ht!]
    \centering
    \includegraphics[width=0.9\textwidth]{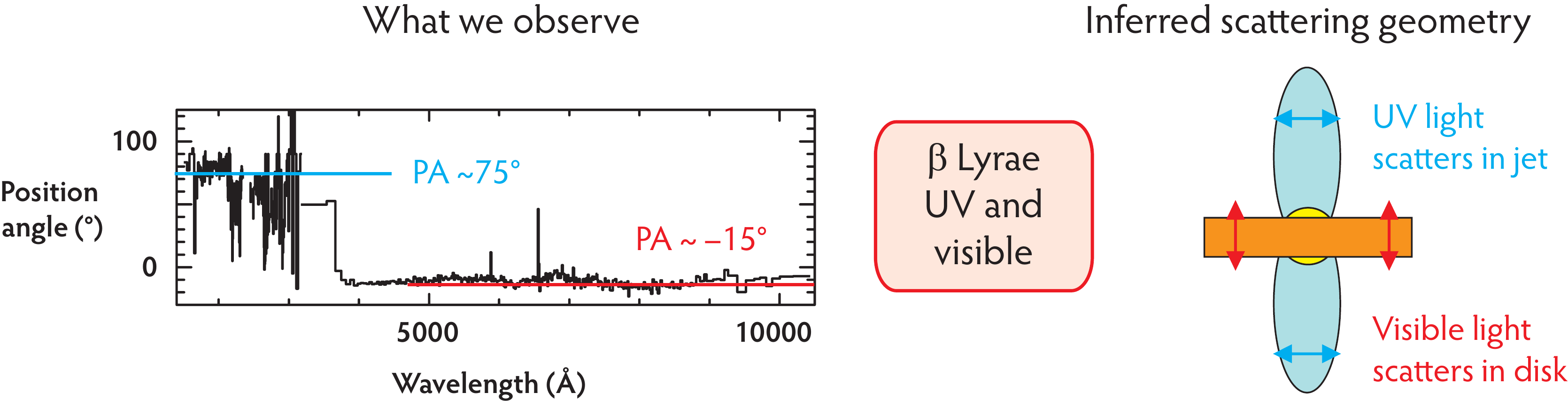}
    \caption{Polarized spectrum (\textit{left}) of $\beta$ Lyr and interpretive picture (\textit{right}), based on the study by \citet{Hoffman1998}. The $90^\circ$ position angle flip between the near-UV and optical continuum provided key evidence for a jet or outflow in the system oriented orthogonal to the thick accretion disk.}
    \label{fig:betlyr}
\end{figure*}

UV polarimetry has proven instrumental in detecting the signatures of mass loss from massive interacting binary systems. In the canonical Algol system $\beta$ Lyr, \citet{Hoffman1998} showed that the near-UV continuum is polarized at $90^\circ$ to the optical continuum, indicating that two orthogonal scattering regions exist in the system. These authors hypothesized that the PA rotation across the Balmer jump occurs because UV light is preferentially absorbed by the thick disk, so is only seen when it scatters in a perpendicular bipolar flow (Figure~\ref{fig:betlyr}). This picture was confirmed by interferometric observations \citep{Harmanec1996,Zhao2008}. By contrast, the similar interacting binary V356 Sgr showed no significant PA difference between the optical and the near-UV \citep{Lomax2017}. This suggests that these Roche-lobe filling binary systems can take on a diversity of geometrical structures, and that using an instrument such as \textit{Polstar} to characterize a broader sample of these objects will allow deeper insights into their physical nature and evolutionary history.

\citet{Lomax2012} used multiwavelength polarimetry to constrain the location and extent of the hot spot on the disk edge in $\beta$ Lyr; extending this analysis into the UV would provide additional 3-dimensional information about the location of hot gas in this and similar systems. In particular, the light curve of $\beta$ Lyr is known to be strongly wavelength-dependent, with significant eclipse reversals at shorter wavelengths \citep{Kondo1994,Ignace2008}, suggesting that time-dependent polarimetric monitoring will open a new, heretofore unexplored window into this and similar systems.

Several of the emission lines in $\beta$ Lyr show polarization position angles that agree with the UV continuum, suggesting that they scatter primarily from the jet or outflow \citep{Hoffman1998}. In the colliding-wind binary system V444 Cyg, the polarized emission lines probe the stratified layers of the WN star's wind and correlate strongly with the wind geometry as inferred from X-ray emission \citep{Lomax2015}. These findings suggest that the time-dependent line polarization data obtained by \textit{Polstar} will provide detailed information about the structure of the winds in our massive binary targets.

\section{Spectroscopic Signatures of Circumstellar Plasma Structures in the FUV}

Circumstellar and photospheric structures that have been identified from FUV spectra are shown in Figure~\ref{fig:ryperfig} and include the gas stream, hot accretion spot, splash plasma, mass loss at phase 0.5 (superior conjunction of the mass loser), jets (mass loss above/below the orbital plane), and a high-temperature plasma \citep{2001ASSL..264...79P,2004AN....325..225P}.

\begin{figure}[b!]
\begin{center}
\includegraphics[width=8.5cm]{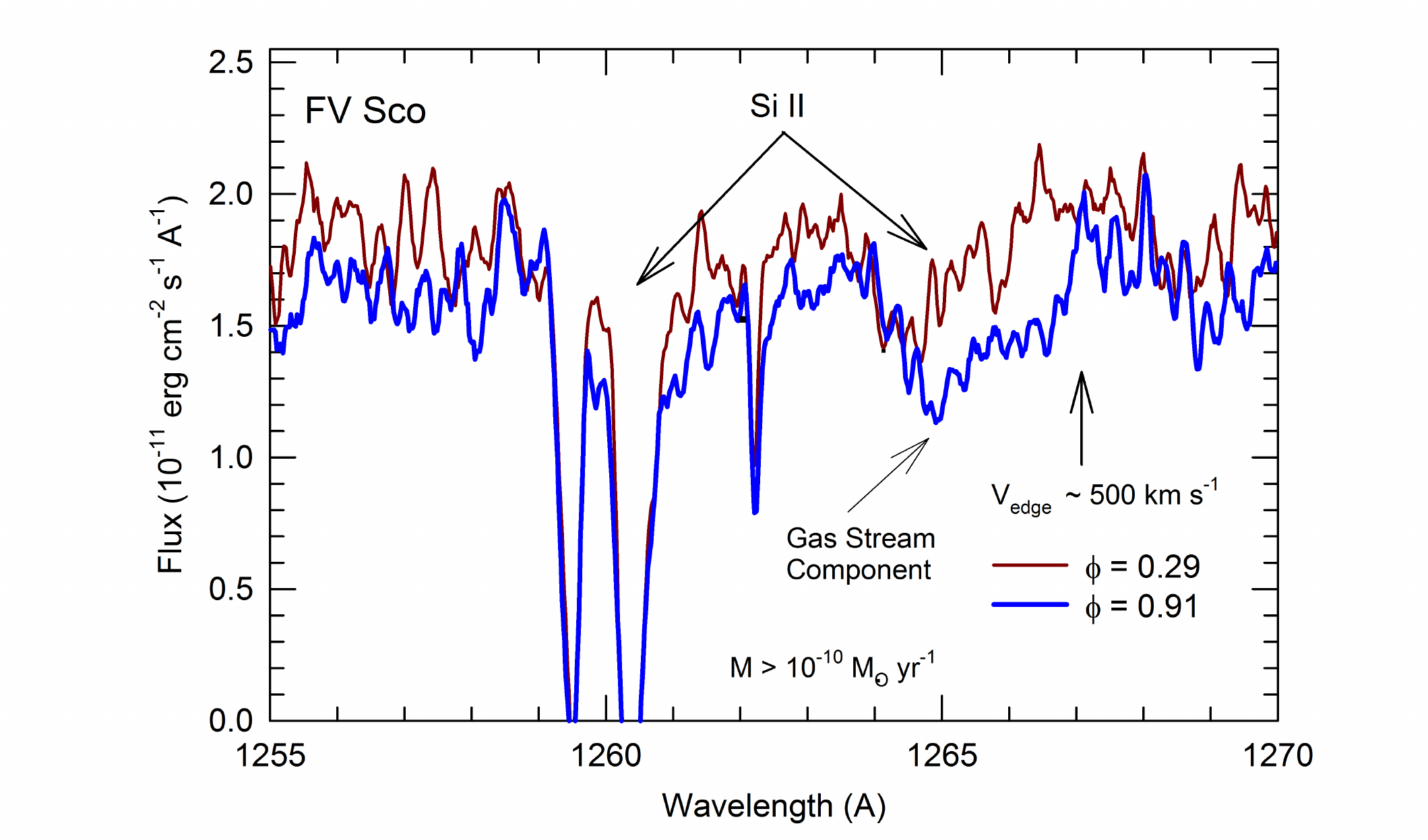}
\caption{The Si~{\sc ii} resonance lines at 126~nm observed at phases 0.91 \& 0.29.  Note the red-shifted gas stream component at phase 0.91.} 
\label{massflowsi2}
\end{center}
\end{figure}

The gas streams in Algol systems are usually detected from {\it additional} 
red-shifted absorption in the moderate-ionization species (e.g. Si~{\sc ii, iii},  
S~{\sc ii, iii},  A$\ell$~{\sc ii, iii}, Mg~{\sc ii}) in the orbital phase interval 
$\rm 0.70 < \phi_{o} < 0.95$, while enhancements in the mass loss are
generally seen from violet-shifted absorption in the Si~{\sc iv} wind lines. An example of gas stream absorption is given in Figure~\ref{massflowsi2}. Gas stream signatures can appear either as discrete absorption or an \textit{inverse} wind feature.

\begin{figure}[h!]
\begin{center}
\includegraphics[width=7.0cm]{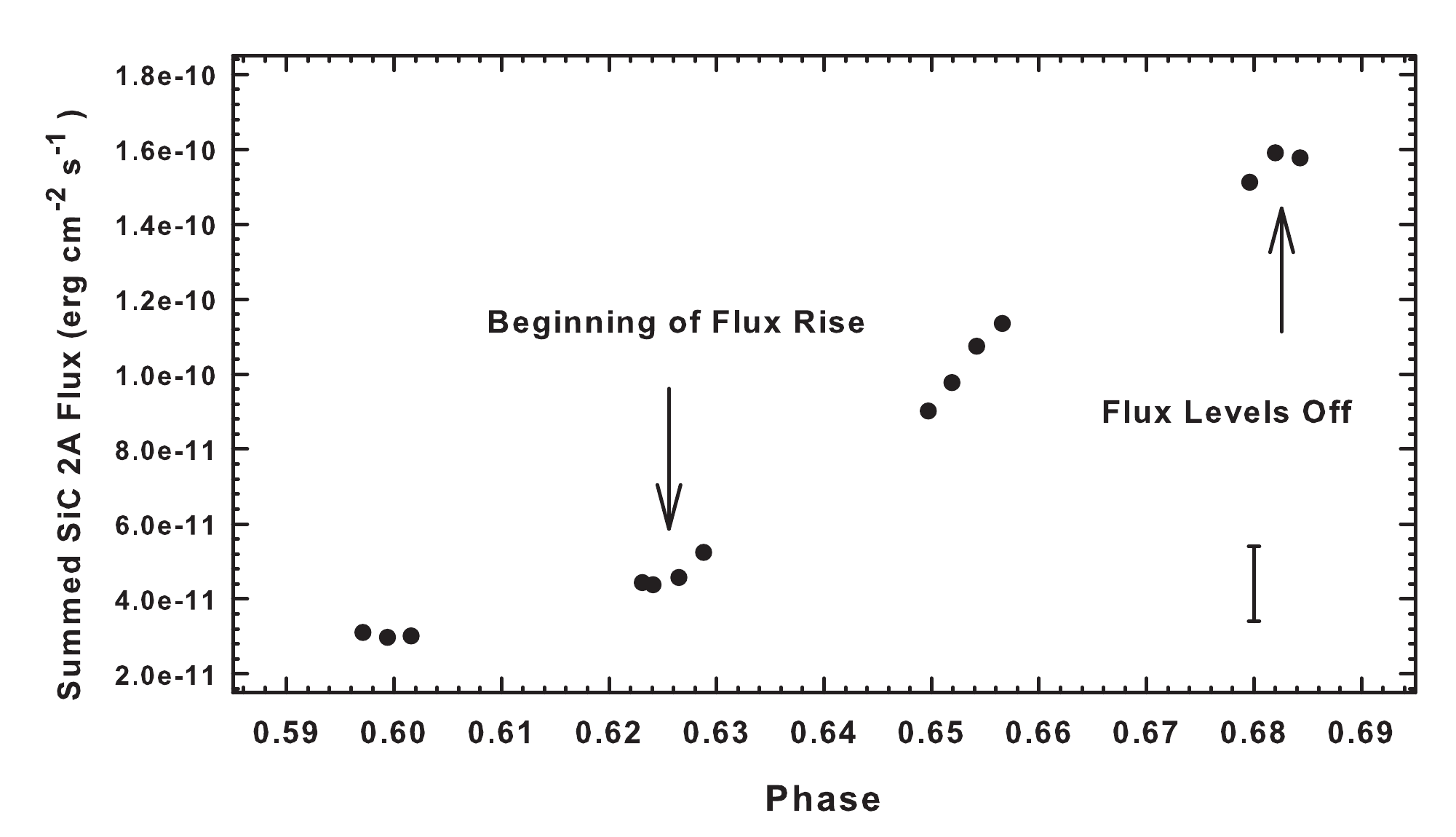}
\includegraphics[width=7.0cm]{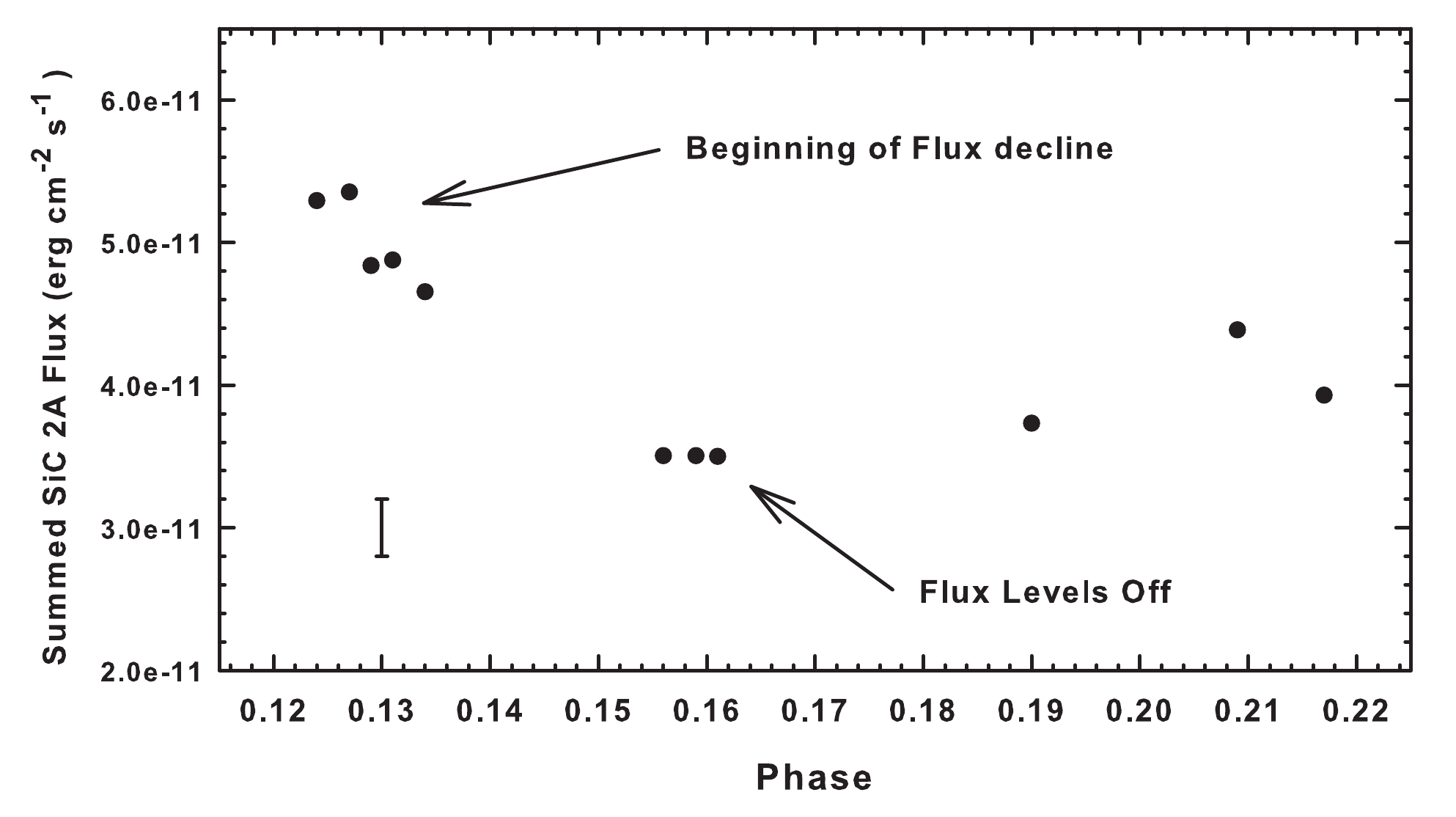}
\caption{Evidence for a hot accretion spot in U Cep from \textit{FUSE} data.
The summed flux from the SiC~2A detector for each exposure obtained 
on 2004 February 09 and 2007 March 10 is plotted versus phase. In the 
\textit{upper} panel note that an apparent hot spot rotates into our line-of-sight at 
phase 0.625 and is fully visible at phase 0.68.  In the \textit{lower} panel we observe this hot spot rotating out of our view a half phase later.}
\label{hotspot}
\end{center}
\end{figure}

Recall that if the system is located in \textit{Region~1} of the \textit{r-q}~diagram, the mass gainer will present a large enough cross-section that the gas stream 
will directly impact its photosphere. Since the impact velocity is $\rm \sim400~km~s^{-1}$,
one expects some degree of shock heating at the impact site.
Observations of the direct impact system U~Cep (P=2\fd49) during the lifetime of the {\it FUSE} spacecraft present convincing evidence for an accretion hot spot \citep{2007IAUS..240..148P}.  
In Figure~\ref{hotspot} we show an apparent hot spot rotating into, then out of our line-of-sight.
Photometric timing allows one to determine the longitude and size of the hot spot.
We estimate that the hot spot is located at the substellar point associated with 0\,\fp90, 
covers about 2\% of the facing photosphere, and has a temperature of $\sim$30000~K which is about three
times the $\rm T_{eff}$ of the mass gainer. We plan to look for an elevated FUV flux versus phase in the other \textit{Polstar} targets listed in Table~1 to identify/map hot impact regions in other Algols.
The role of hot accretion 
spots in binary star evolution is discussed by \citet{2008A&A...487.1129V,2010A&A...510A..13V} who concluded that such spots 
can help drive mass out of the system and significantly affect the final products of the mass transfer. A likely place to find this type of mass loss would be above/below the orbital plane.

\begin{figure}[h!]
\begin{center}
\includegraphics[width=8.0cm]{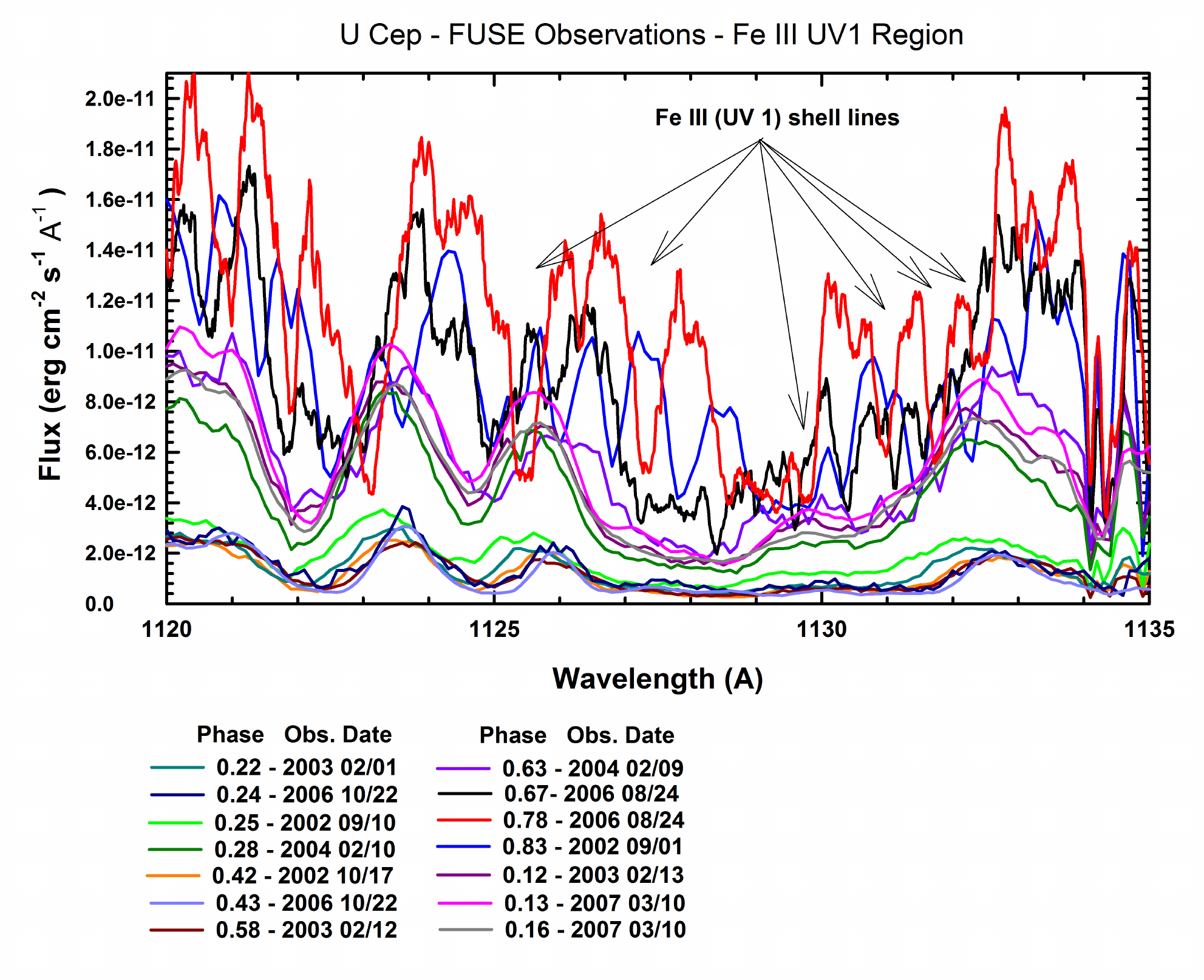}
\caption{FUSE spectra of U Cep. Note the elevated flux at 
phases 0.63, 0.67, 0.78, 0.83, 0.12, 0.13, and 0.16.  Obvious "shell" structure is observed at phases 0.66, 0.78, and 0.83, but none at the post-conjunction phases even though the hot spot is still visible. The former suggest outflowing plasma from a splash, region while the latter one reveals infall of material from the gas stream.}
\label{ucepsplash}
\end{center}
\end{figure}

Associated with the accretion spot in U~Cep  is a localized circumstellar plasma that hovers
over the impact region.  It is identified from the
phase-dependence of the strengths of shell lines of moderate-ionization species. The
behavior is shown in Figure~\ref{ucepsplash} in which spectra centered on the Fe~{\sc iii} (UV1) resonance multiplet are compared. \textit{Shell-type} absorption features formed
in this plasma are observed at phases 0.66, 0.78, and 0.83 but not at
phases 0.12, 0.13, and 0.16 even though the hot accretion spot is still visible
on the \textit{receding} limb and the FUV flux elevated. Only photospheric lines are
observed. When the hot spot is not visible, a pure
photospheric spectrum is observed. The localized circumstellar plasma is probably
formed as the result of a \textit{splash} associated with the impacting gas stream.
The measured velocities of the Fe~{\sc iii} shell lines are blue-shifted relative to
the photosphere before phase 0.66 but red-shifted in the interval 0.78-0.83.
The shell lines from the latter phase interval are apparently formed in the
gas stream. Since Fe~{\sc iii} and similar species appear to be dominant,
the temperature in the plasma is in the range of 20-40~ kK. A splash plasma has also been identified in RY~Per \citep{2004ApJ...608..989B}. Hot accretion spots and splash plasmas may be commonplace in Algol binaries. Now that we have observational confirmation of their existence, \textit{Polstar} is poised to yield the data that will allow us to map out the angular extent of such structures and determine how much of the splashed material actually escapes from the system.

\begin{figure}[h!]
\begin{center}
\includegraphics[width=8.5cm]{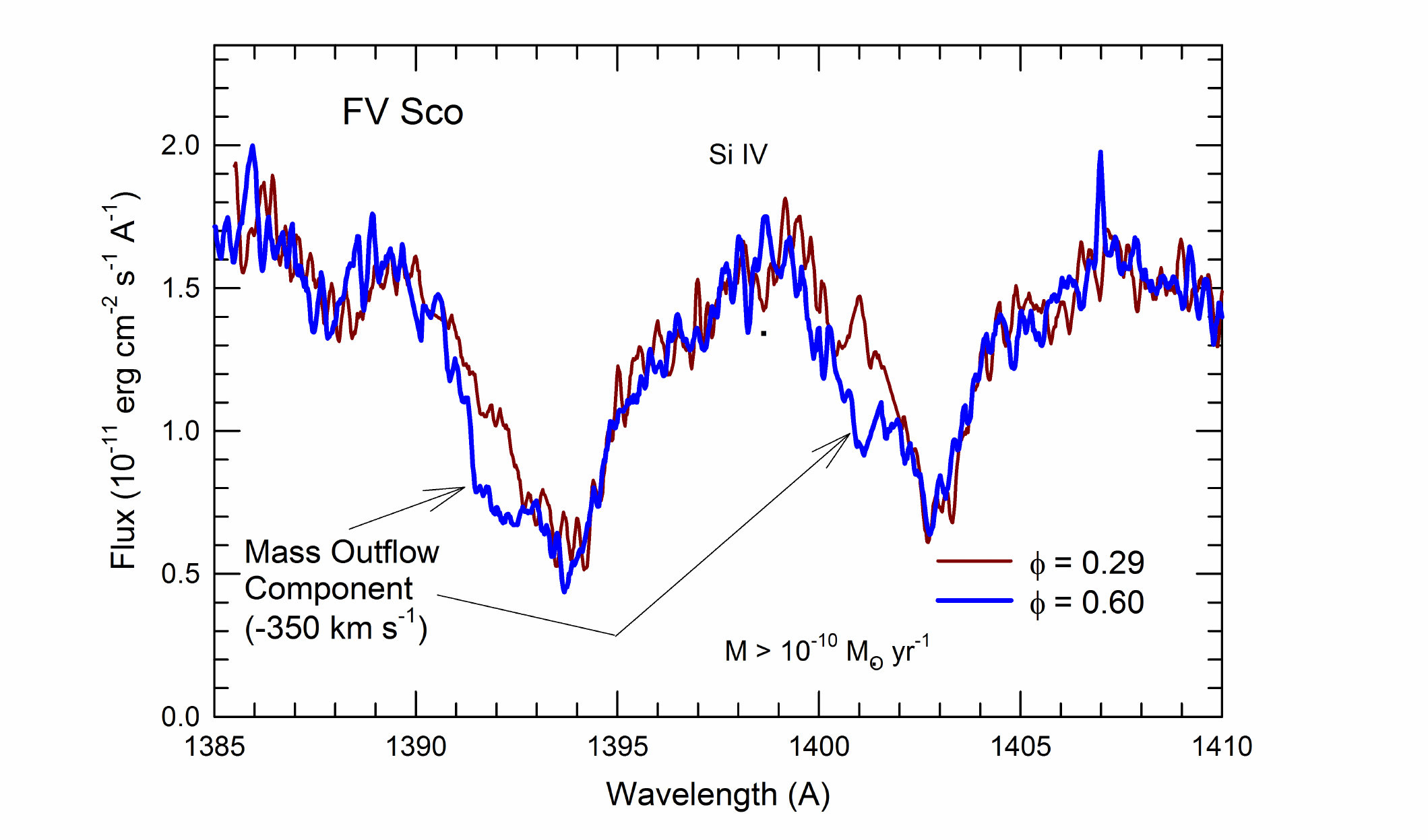}
\caption{The Si~{\sc iv} resonance lines at 140~nm observed at phases 0.91 \& 0.29.  
Evidence for extensive mass loss at phase 0.60 is seen from a comparison of the Si~{\sc iv} resonance doublet observed at this phase with a similar observation near quadrature.}
\label{massflowsi4}
\end{center}
\end{figure}

Optical and UV studies of Algols have revealed that mass loss near 
$\rm \phi_{o}\sim0.5$ is commonplace \citet{1989SSRv...50....9P,2001ASSL..264...79P}. Often times it is obvious that the flow is a collimated structure caused by the deflection of the gas stream as it impacts the mass gainer's photosphere (cf. Figure~\ref{massflowsi4}) at a shallow angle. 
The mass and angular momentum loss from an Algol system is important for 
theoretical modeling of its evolution, but this quantity, which remains one of the least known, can be modeled from \textit{Polstar} spectroscopic data through spectrum synthesis techniques. Evolution models with nonconservative mass transfer appear to produce better matches to the observed stellar parameters \citep[e.g.,][]{2008A&A...487.1129V,2010A&A...510A..13V,2001ApJ...552..664N}. But this is not a new conclusion. 
About 40 years ago statistical evidence for {\it extensive} mass loss during the mass transfer was presented by \citet{1981ApJS...46....1G} and \citet{1983ApJS...52...35G}.
These authors considered several conservative and nonconservative studies
and concluded that mass and angular momentum loss occurred in 20\% of their
sample.

Jets, or more generally regions of mass loss above/below the orbital plane, may be commonplace in Algol binaries, but they are difficult to detect because most of their material is not along our line-of-sight to the mass gainer. There are both polarimetric  \citep[e.g.,][]{Hoffman1998} and FUV spectroscopic \citep[e.g.,][]{2004AN....325..225P,1980IAUS...88..251P} signatures.  Emission lines are often observed during a total eclipse of the mass gainer. These (usually broad) features are from species of moderate-high ionization. Noteworthy are the O~{\sc vi} and N~{\sc v} resonance doublets that indicate plasma temperatures of 300--100~kK.

The discovery of prominent absorption and emission lines of N~{\sc v}, C~{\sc iv}, and Si~{\sc iv} in the UV spectra of Algol-type binaries in the early 1980s with \textit{IUE} revealed the presence of a high temperature component to the circumstellar material  \citep{1984ApJ...283..745P,1980IAUS...88..251P,1983ApJ...275..251P}.
Peters and Polidan called this circumstellar structure the High Temperature Accretion Region (HTAR). They concluded that the HTAR (absorption) lines 
are formed by resonance scattering in a region of $\rm T_{e}\sim10^{5}$~K,
$\rm N_{e}\sim10^{9}~cm^{-3}$, and extreme carbon depletion,
$\rm C\sim10^{-3}~C_{\odot}$. The radial velocity behavior of the UV
absorption lines provide compelling evidence that the high temperature plasma
is associated with the mass gainer.  Since the high temperature plasma
appears to be mostly detected on the {\it trailing} hemisphere of the primary,
it was concluded that the plasma producing the HTAR absorption lines is most likely
heated by a shock that occurs as the gas stream impacts either the
primary's photosphere or inner accretion disk. In some of the systems the 
material currently being transferred appears to have been processed through the 
CNO-cycle in the core of the mass loser. Phase-resolved FUV spectroscopy with \textit{Polstar} will allow us to map the HTAR with more precision and determine if it is an important source of systemic mass loss.

\section{Concluding Remarks}
In this paper we have discussed how we can quantify the amount of systemic mass loss in OB close binaries from FUV polarimetry and high-resolution spectroscopy with \textit{Polstar}. Since 50-100\% of the OB stars are formed with one or more companions, stellar evolution calculations for the upper main sequence must consider mass transfer in close binary systems. We have identified two stages in interacting binary evolution that are important epochs of systemic mass and angular momentum loss:  prior to mass reversal and when the system passes through \textit{Region 2} in the \textit{r-q}~diagram after reversal. We have identified 24 targets that represent all three regions in the \textit{r-q}~diagram and also have known parameters for their mass gainers and losers as well as the geometry, sizes, and approximate physical conditions in one or more circumstellar structures. With 20 observation pairs (spectroscopy and polarimetry) we will achieve complete phase-coverage with more observations taking place around the eclipse of the mass gainer where mapping of the asymmetrical circumstellar structures will be of high precision. In Section 3 we discussed the observing program in detail and show how we can measure the mass flow and loss in an OB interacting binary with \textit{Polstar} with the necessary signal-to-noise and precision that is needed to quantify the amount of nonconservative mass transfer.  Combined with the objectives in \textit{S3} (Jones et al 2022, this Topical Collection) which  will address the binary fraction and angular momentum issues, this project will lead to improved evolutionary tracks for OB stars and a more realistic view of their post-main sequence evolution.


\bibliography{S4}

\begin{thebibliography}{65}
\providecommand{\natexlab}[1]{#1}
\providecommand{\url}[1]{{#1}}
\providecommand{\urlprefix}{URL }
\providecommand{\doi}[1]{\url{https://doi.org/#1}}
\providecommand{\eprint}[2][]{\url{#2}}
 \bibcommenthead

\bibitem[{{Astropy Collaboration} et~al(2013){Astropy Collaboration},
  {Robitaille}, {Tollerud}, {Greenfield}, {Droettboom}, {Bray}, {Aldcroft},
  {Davis}, {Ginsburg}, {Price-Whelan}, {Kerzendorf}, {Conley}, {Crighton},
  {Barbary}, {Muna}, {Ferguson}, {Grollier}, {Parikh}, {Nair}, {Unther},
  {Deil}, {Woillez}, {Conseil}, {Kramer}, {Turner}, {Singer}, {Fox}, {Weaver},
  {Zabalza}, {Edwards}, {Azalee Bostroem}, {Burke}, {Casey}, {Crawford},
  {Dencheva}, {Ely}, {Jenness}, {Labrie}, {Lim}, {Pierfederici}, {Pontzen},
  {Ptak}, {Refsdal}, {Servillat}, and {Streicher}}]{astropy2013}
{Astropy Collaboration}, {Robitaille} TP, {Tollerud} EJ, et~al (2013) {Astropy:
  A community Python package for astronomy}. \aap 558:A33.
  \doi{10.1051/0004-6361/201322068},
  {\href{https://arxiv.org/abs/1307.6212}{{https://arxiv.org/abs/arXiv:1307.6212}}}
  {[astro-ph.IM]}

\bibitem[{{Astropy Collaboration} et~al(2018){Astropy Collaboration},
  {Price-Whelan}, {Sip{\H{o}}cz}, {G{\"u}nther}, {Lim}, {Crawford}, {Conseil},
  {Shupe}, {Craig}, {Dencheva}, {Ginsburg}, {Vand erPlas}, {Bradley},
  {P{\'e}rez-Su{\'a}rez}, {de Val-Borro}, {Aldcroft}, {Cruz}, {Robitaille},
  {Tollerud}, {Ardelean}, {Babej}, {Bach}, {Bachetti}, {Bakanov}, {Bamford},
  {Barentsen}, {Barmby}, {Baumbach}, {Berry}, {Biscani}, {Boquien}, {Bostroem},
  {Bouma}, {Brammer}, {Bray}, {Breytenbach}, {Buddelmeijer}, {Burke},
  {Calderone}, {Cano Rodr{\'\i}guez}, {Cara}, {Cardoso}, {Cheedella}, {Copin},
  {Corrales}, {Crichton}, {D'Avella}, {Deil}, {Depagne}, {Dietrich}, {Donath},
  {Droettboom}, {Earl}, {Erben}, {Fabbro}, {Ferreira}, {Finethy}, {Fox},
  {Garrison}, {Gibbons}, {Goldstein}, {Gommers}, {Greco}, {Greenfield},
  {Groener}, {Grollier}, {Hagen}, {Hirst}, {Homeier}, {Horton}, {Hosseinzadeh},
  {Hu}, {Hunkeler}, {Ivezi{\'c}}, {Jain}, {Jenness}, {Kanarek}, {Kendrew},
  {Kern}, {Kerzendorf}, {Khvalko}, {King}, {Kirkby}, {Kulkarni}, {Kumar},
  {Lee}, {Lenz}, {Littlefair}, {Ma}, {Macleod}, {Mastropietro}, {McCully},
  {Montagnac}, {Morris}, {Mueller}, {Mumford}, {Muna}, {Murphy}, {Nelson},
  {Nguyen}, {Ninan}, {N{\"o}the}, {Ogaz}, {Oh}, {Parejko}, {Parley}, {Pascual},
  {Patil}, {Patil}, {Plunkett}, {Prochaska}, {Rastogi}, {Reddy Janga},
  {Sabater}, {Sakurikar}, {Seifert}, {Sherbert}, {Sherwood-Taylor}, {Shih},
  {Sick}, {Silbiger}, {Singanamalla}, {Singer}, {Sladen}, {Sooley},
  {Sornarajah}, {Streicher}, {Teuben}, {Thomas}, {Tremblay}, {Turner},
  {Terr{\'o}n}, {van Kerkwijk}, {de la Vega}, {Watkins}, {Weaver}, {Whitmore},
  {Woillez}, {Zabalza}, and {Astropy Contributors}}]{astropy2018}
{Astropy Collaboration}, {Price-Whelan} AM, {Sip{\H{o}}cz} BM, et~al (2018)
  {The Astropy Project: Building an Open-science Project and Status of the v2.0
  Core Package}. \aj 156(3):123. \doi{10.3847/1538-3881/aabc4f},
  {\href{https://arxiv.org/abs/1801.02634}{{https://arxiv.org/abs/arXiv:1801.02634}}}
  {[astro-ph.IM]}

\bibitem[{{Barai} et~al(2004){Barai}, {Gies}, {Choi}, {Das}, {Deo}, {Huang},
  {Marshall}, {McSwain}, {Ogden}, {Osterman}, {Riddle}, {Seymour}, {Wingert},
  {Kaye}, and {Peters}}]{2004ApJ...608..989B}
{Barai} P, {Gies} DR, {Choi} E, et~al (2004) {Mass and Angular Momentum
  Transfer in the Massive Algol Binary RY Persei}. \apj 608(2):989--1000.
  \doi{10.1086/420875},
  {\href{https://arxiv.org/abs/astro-ph/0309734}{{https://arxiv.org/abs/arXiv:astro-ph/0309734}}}
  {[astro-ph]}

\bibitem[{{Brown} et~al(1978){Brown}, {McLean}, and
  {Emslie}}]{1978A&A....68..415B}
{Brown} JC, {McLean} IS, {Emslie} AG (1978) {Polarisation by Thomas scattering
  in optically thin stellar envelopes. II. Binary and multiple star envelopes
  and the determination of binary inclinations.} \aap 68:415--427

\bibitem[{{Brown} et~al(1982){Brown}, {Aspin}, {Simmons}, and
  {McLean}}]{1982MNRAS.198..787B}
{Brown} JC, {Aspin} C, {Simmons} JFL, et~al (1982) {The effect of orbital
  eccentricity on polarimetric binary diagnostics.} \mnras 198:787--794.
  \doi{10.1093/mnras/198.3.787}

\bibitem[{{Carciofi} et~al(2017){Carciofi}, {Bjorkman}, and
  {Zsarg{\'o}}}]{2017IAUS..329..390C}
{Carciofi} AC, {Bjorkman} JE, {Zsarg{\'o}} J (2017) {HDUST3 - A chemically
  realistic, 3-D, NLTE radiative transfer code}. In: {Eldridge} JJ, {Bray} JC,
  {McClelland} LAS, et~al (eds) The Lives and Death-Throes of Massive Stars, pp
  390--390, \doi{10.1017/S1743921317002836}

\bibitem[{{Clayton} et~al(1992){Clayton}, {Anderson}, {Magalhaes}, {Code},
  {Nordsieck}, {Meade}, {Wolff}, {Babler}, {Bjorkman}, {Schulte-Ladbeck},
  {Taylor}, and {Whitney}}]{1992ApJ...385L..53C}
{Clayton} GC, {Anderson} CM, {Magalhaes} AM, et~al (1992) {The First
  Spectropolarimetric Study of the Wavelength Dependence of Interstellar
  Polarization in the Ultraviolet}. \apjl 385:L53. \doi{10.1086/186276}

\bibitem[{{Deschamps} et~al(2015){Deschamps}, {Braun}, {Jorissen}, {Siess},
  {Baes}, and {Camps}}]{2015A&A...577A..55D}
{Deschamps} R, {Braun} K, {Jorissen} A, et~al (2015) {Non-conservative
  evolution in Algols: where is the matter?} \aap 577:A55.
  \doi{10.1051/0004-6361/201424772},
  {\href{https://arxiv.org/abs/1502.04957}{{https://arxiv.org/abs/arXiv:1502.04957}}}
  {[astro-ph.SR]}

\bibitem[{{Drechsel} and {Lorenz}(1993)}]{1993IBVS.3868....1D}
{Drechsel} H, {Lorenz} R (1993) {Period of SV Centauri Continues Decreasing}.
  Information Bulletin on Variable Stars 3868:1

\bibitem[{{Drechsel} et~al(1982){Drechsel}, {Rahe}, {Wargau}, and
  {Wolf}}]{1982A&A...110..246D}
{Drechsel} H, {Rahe} J, {Wargau} W, et~al (1982) {The interacting early-type
  contact binary SV Cen.} \aap 110:246--262

\bibitem[{Eldridge et~al(2013)Eldridge, Fraser, Smartt, Maund, and
  Crockett}]{eldridge2013}
Eldridge JJ, Fraser M, Smartt SJ, et~al (2013) The death of massive stars -
  {{II}}. {{Observational}} constraints on the progenitors of {{Type Ibc}}
  supernovae. Monthly Notices of the Royal Astronomical Society 436:774--795.
  \doi{10.1093/mnras/stt1612}

\bibitem[{{Fox}(1991)}]{1991ApJ...379..663F}
{Fox} GK (1991) {Stellar Occultation of Polarized Light from Circumstellar
  Electrons. III. General Axisymmetric Envelopes}. \apj 379:663.
  \doi{10.1086/170540}

\bibitem[{{Fox}(1994)}]{1994ApJ...435..372F}
{Fox} GK (1994) {The Theoretical Polarization of Pure Scattering Axisymmetric
  Circumstellar Envelopes}. \apj 435:372. \doi{10.1086/174819}

\bibitem[{{Giuricin} and {Mardirossian}(1981)}]{1981ApJS...46....1G}
{Giuricin} G, {Mardirossian} F (1981) {Some aspects of mass loss and mass
  transfer in Algol variables.} \apjs 46:1--26. \doi{10.1086/190732}

\bibitem[{{Giuricin} et~al(1983){Giuricin}, {Mardirossian}, and
  {Mezzetti}}]{1983ApJS...52...35G}
{Giuricin} G, {Mardirossian} F, {Mezzetti} M (1983) {General properties of
  Algol binaries.} \apjs 52:35--60. \doi{10.1086/190858}

\bibitem[{{Harmanec} et~al(1996){Harmanec}, {Morand}, {Bonneau}, {Jiang},
  {Yang}, {Guinan}, {Hall}, {Mourard}, {Hadrava}, {Bozic}, {Sterken},
  {Tallon-Bosc}, {Walker}, {McCook}, {Vakili}, {Stee}, and {Le
  Contel}}]{Harmanec1996}
{Harmanec} P, {Morand} F, {Bonneau} D, et~al (1996) {Jet-like structures in
  {\ensuremath{\beta}} Lyrae. Results of optical interferometry, spectroscopy
  and photometry.} \aap 312:879--896

\bibitem[{{Hoffman} et~al(1998){Hoffman}, {Nordsieck}, and {Fox}}]{Hoffman1998}
{Hoffman} JL, {Nordsieck} KH, {Fox} GK (1998) {Spectropolarimetric Evidence for
  a Bipolar Flow in beta Lyrae}. \aj 115(4):1576--1591. \doi{10.1086/300274}

\bibitem[{{Hoffman} et~al(2003){Hoffman}, {Whitney}, and
  {Nordsieck}}]{Hoffman2003}
{Hoffman} JL, {Whitney} BA, {Nordsieck} KH (2003) {The Effect of Multiple
  Scattering on the Polarization from Binary Star Envelopes. I. Self- and
  Externally Illuminated Disks}. \apj 598(1):572--587. \doi{10.1086/378770},
  {\href{https://arxiv.org/abs/astro-ph/0307261}{{https://arxiv.org/abs/arXiv:astro-ph/0307261}}}
  {[astro-ph]}

\bibitem[{{Ignace} et~al(2008){Ignace}, {Oskinova}, {Waldron}, {Hoffman}, and
  {Hamann}}]{Ignace2008}
{Ignace} R, {Oskinova} LM, {Waldron} WL, et~al (2008) {Phase-dependent X-ray
  observations of the {\ensuremath{\beta}} Lyrae system. No eclipse in the soft
  band}. \aap 477(3):L37--L40. \doi{10.1051/0004-6361:20078871},
  {\href{https://arxiv.org/abs/0711.3954}{{https://arxiv.org/abs/arXiv:0711.3954}}}
  {[astro-ph]}

\bibitem[{{Ignace} et~al(2022){Ignace}, {Fullard}, {Shrestha}, {Naz{\'e}},
  {Gayley}, {Hoffman}, {Lomax}, and {St-Louis}}]{2022ApJ...933....5I}
{Ignace} R, {Fullard} A, {Shrestha} M, et~al (2022) {Modeling the Optical to
  Ultraviolet Polarimetric Variability from Thomson Scattering in
  Colliding-wind Binaries}. \apj 933(1):5. \doi{10.3847/1538-4357/ac6fce},
  {\href{https://arxiv.org/abs/2205.07612}{{https://arxiv.org/abs/arXiv:2205.07612}}}
  {[astro-ph.SR]}

\bibitem[{{Ivanova} et~al(2013){Ivanova}, {Justham}, {Chen}, {De Marco},
  {Fryer}, {Gaburov}, {Ge}, {Glebbeek}, {Han}, {Li}, {Lu}, {Marsh},
  {Podsiadlowski}, {Potter}, {Soker}, {Taam}, {Tauris}, {van den Heuvel}, and
  {Webbink}}]{2013A&ARv..21...59I}
{Ivanova} N, {Justham} S, {Chen} X, et~al (2013) {Common envelope evolution:
  where we stand and how we can move forward}. \aapr 21:59.
  \doi{10.1007/s00159-013-0059-2},
  {\href{https://arxiv.org/abs/1209.4302}{{https://arxiv.org/abs/arXiv:1209.4302}}}
  {[astro-ph.HE]}

\bibitem[{Jones et~al(2022)Jones, Labadie-Bartz, Cotton, Naz\'e, Peters,
  Hillier, Neiner, Richardson, Hoffman, Carciofi, Wisniewski, Gayley, Suffak,
  Ignace, and Scowen}]{JonesTC}
Jones CE, Labadie-Bartz J, Cotton DV, et~al (2022) {Ultraviolet
  Spectropolarimetry: on the origin of rapidly rotating B stars}. Ap\&SS
  367:{\bf Topical Collection}

\bibitem[{{Kondo} et~al(1994){Kondo}, {McCluskey}, {Silvis}, {Polidan},
  {McCluskey}, and {Eaton}}]{Kondo1994}
{Kondo} Y, {McCluskey} GE, {Silvis} JMS, et~al (1994) {Ultraviolet Light Curves
  of beta Lyrae: Comparison of OAO A-2, IUE, and Voyager Observations}. \apj
  421:787. \doi{10.1086/173691}

\bibitem[{{Kriz} and {Harmanec}(1975)}]{1975BAICz..26...65K}
{Kriz} S, {Harmanec} P (1975) {A Hypothesis of the Binary Origin of Be Stars}.
  Bulletin of the Astronomical Institutes of Czechoslovakia 26:65

\bibitem[{{Langer}(2012)}]{Langer2012}
{Langer} N (2012) {Presupernova Evolution of Massive Single and Binary Stars}.
  \araa 50:107--164. \doi{10.1146/annurev-astro-081811-125534},
  {\href{https://arxiv.org/abs/1206.5443}{{https://arxiv.org/abs/arXiv:1206.5443}}}
  {[astro-ph.SR]}

\bibitem[{{Laplace} et~al(2021){Laplace}, {Justham}, {Renzo}, {G{\"o}tberg},
  {Farmer}, {Vartanyan}, and {de Mink}}]{Laplace2021}
{Laplace} E, {Justham} S, {Renzo} M, et~al (2021) {Different to the core: the
  pre-supernova structures of massive single and binary-stripped stars}. arXiv
  e-prints arXiv:2102.05036.
  {\href{https://arxiv.org/abs/2102.05036}{{https://arxiv.org/abs/arXiv:2102.05036}}}
  {[astro-ph.SR]}

\bibitem[{{Linnell} and {Scheick}(1991)}]{1991ApJ...379..721L}
{Linnell} AP, {Scheick} X (1991) {Does SV Centauri Harbor an Accretion Disk?}
  \apj 379:721. \doi{10.1086/170547}

\bibitem[{{Linnell} et~al(1988){Linnell}, {Peters}, and
  {Polidan}}]{1988ApJ...327..265L}
{Linnell} AP, {Peters} GJ, {Polidan} RS (1988) {An Improved Photometric
  Analysis of SX Aurigae}. \apj 327:265. \doi{10.1086/166187}

\bibitem[{{Lomax} et~al(2012{\natexlab{a}}){Lomax}, {Hoffman}, {Elias},
  {Bastien}, and {Holenstein}}]{Lomax2012}
{Lomax} JR, {Hoffman} JL, {Elias} INicholas~M., et~al (2012{\natexlab{a}})
  {Geometrical Constraints on the Hot Spot in Beta Lyrae}. \apj 750(1):59.
  \doi{10.1088/0004-637X/750/1/59},
  {\href{https://arxiv.org/abs/1108.3015}{{https://arxiv.org/abs/arXiv:1108.3015}}}
  {[astro-ph.SR]}

\bibitem[{{Lomax} et~al(2012{\natexlab{b}}){Lomax}, {Hoffman}, {Elias},
  {Bastien}, and {Holenstein}}]{2012ApJ...750...59L}
{Lomax} JR, {Hoffman} JL, {Elias} INicholas~M., et~al (2012{\natexlab{b}})
  {Geometrical Constraints on the Hot Spot in Beta Lyrae}. \apj 750(1):59.
  \doi{10.1088/0004-637X/750/1/59},
  {\href{https://arxiv.org/abs/1108.3015}{{https://arxiv.org/abs/arXiv:1108.3015}}}
  {[astro-ph.SR]}

\bibitem[{{Lomax} et~al(2015){Lomax}, {Naz{\'e}}, {Hoffman}, {Russell}, {De
  Becker}, {Corcoran}, {Davidson}, {Neilson}, {Owocki}, {Pittard}, and
  {Pollock}}]{Lomax2015}
{Lomax} JR, {Naz{\'e}} Y, {Hoffman} JL, et~al (2015) {V444 Cygni X-ray and
  polarimetric variability: Radiative and Coriolis forces shape the wind
  collision region}. \aap 573:A43. \doi{10.1051/0004-6361/201424468},
  {\href{https://arxiv.org/abs/1410.6117}{{https://arxiv.org/abs/arXiv:1410.6117}}}
  {[astro-ph.SR]}

\bibitem[{{Lomax} et~al(2017){Lomax}, {Fullard}, {Malatesta}, {Babler},
  {Bednarski}, {Berdis}, {Bjorkman}, {Bjorkman}, {Carciofi}, {Davidson},
  {Keil}, {Meade}, {Nordsieck}, {Scheffler}, {Hoffman}, and
  {Wisniewski}}]{Lomax2017}
{Lomax} JR, {Fullard} AG, {Malatesta} MA, et~al (2017) {The complex
  circumstellar and circumbinary environment of V356 Sgr}. \mnras
  464(2):1936--1947. \doi{10.1093/mnras/stw2457},
  {\href{https://arxiv.org/abs/1609.07489}{{https://arxiv.org/abs/arXiv:1609.07489}}}
  {[astro-ph.SR]}

\bibitem[{{Lubow} and {Shu}(1975)}]{1975ApJ...198..383L}
{Lubow} SH, {Shu} FH (1975) {Gas dynamics of semidetached binaries.} \apj
  198:383--405. \doi{10.1086/153614}

\bibitem[{{Lubow} and {Shu}(1976)}]{1976ApJ...207L..53L}
{Lubow} SH, {Shu} FH (1976) {Gas dynamics of semidetached binaries. II. The
  vertical structure of the stream.} \apjl 207:L53--L55. \doi{10.1086/182177}

\bibitem[{{Manset} and {Bastien}(2000)}]{2000AJ....120..413M}
{Manset} N, {Bastien} P (2000) {Polarimetric Variations of Binary Stars. I.
  Numerical Simulations for Circular and Eccentric Binaries in Thomson
  Scattering Envelopes}. \aj 120(1):413--429. \doi{10.1086/301439}

\bibitem[{{Nelson} and {Eggleton}(2001)}]{2001ApJ...552..664N}
{Nelson} CA, {Eggleton} PP (2001) {A Complete Survey of Case A Binary Evolution
  with Comparison to Observed Algol-type Systems}. \apj 552(2):664--678.
  \doi{10.1086/320560},
  {\href{https://arxiv.org/abs/astro-ph/0009258}{{https://arxiv.org/abs/arXiv:astro-ph/0009258}}}
  {[astro-ph]}

\bibitem[{{Nordsieck} et~al(2001){Nordsieck}, {Wisniewski}, {Babler}, {Meade},
  {Anderson}, {Bjorkman}, {Code}, {Fox}, {Johnson}, {Weitenbeck}, and
  {Zellner}}]{2001ASPC..233..261N}
{Nordsieck} KH, {Wisniewski} J, {Babler} BL, et~al (2001) {Ultraviolet and
  visible spectropolarimetric variability in P Cygni}. In: {de Groot} M,
  {Sterken} C (eds) P Cygni 2000: 400 Years of Progress, p 261,
  \eprint{astro-ph/0102073}

\bibitem[{{Paczy{\'n}ski}(1971)}]{Paczynski1971}
{Paczy{\'n}ski} B (1971) {Evolutionary Processes in Close Binary Systems}.
  \araa 9:183. \doi{10.1146/annurev.aa.09.090171.001151}

\bibitem[{{Peters}(1989)}]{1989SSRv...50....9P}
{Peters} GJ (1989) {The H{\ensuremath{\alpha}} Emitting Regions of the
  Accretion Disks in ALGOLS}. \ssr 50(1-2):9--22. \doi{10.1007/BF00215915}

\bibitem[{{Peters}(2001)}]{2001ASSL..264...79P}
{Peters} GJ (2001) {The Algol-Type Binaries}. In: {Vanbeveren} D (ed) The
  Influence of Binaries on Stellar Population Studies, p~79,
  \doi{10.1007/978-94-015-9723-4\_6}

\bibitem[{{Peters}(2007)}]{2007IAUS..240..148P}
{Peters} GJ (2007) {Bipolar Jets, Hot Interaction Regions, and Colliding Winds
  in OB Interacting Binaries}. In: {Hartkopf} WI, {Harmanec} P, {Guinan} EF
  (eds) Binary Stars as Critical Tools \& Tests in Contemporary Astrophysics,
  pp 148--153, \doi{10.1017/S174392130700395X}

\bibitem[{{Peters} and {Polidan}(1984)}]{1984ApJ...283..745P}
{Peters} GJ, {Polidan} RS (1984) {Evidence for a high-temperature accretion
  region in Algol-type binarysystems.} \apj 283:745--759. \doi{10.1086/162359}

\bibitem[{{Peters} and {Polidan}(2004)}]{2004AN....325..225P}
{Peters} GJ, {Polidan} RS (2004) {Eclipse mapping of the hot circumstellar
  plasma in Algol binaries}. Astronomische Nachrichten 325(3):225--228.
  \doi{10.1002/asna.200310224}

\bibitem[{{Plavec}(1970)}]{1970PASP...82..957P}
{Plavec} M (1970) {Mass Exchange in Binary Stars}. \pasp 82(489):957.
  \doi{10.1086/128996}

\bibitem[{{Plavec}(1976)}]{1976IAUS...70..439P}
{Plavec} M (1976) {Final Remarks on the Binary Hypothesis for the be Stars}.
  In: {Slettebak} A (ed) Be and Shell Stars, p 439

\bibitem[{{Plavec} and {Polidan}(1976)}]{1976IAUS...73..289P}
{Plavec} M, {Polidan} RS (1976) {The ALGOLS, Red Spectra, BE Stars, and Even
  Neutrinos}. In: {Eggleton} P, {Mitton} S, {Whelan} J (eds) Structure and
  Evolution of Close Binary Systems, p 289

\bibitem[{{Plavec}(1980)}]{1980IAUS...88..251P}
{Plavec} MJ (1980) {IUE observations of long period eclipsing binaries: A study
  of accretion onto non-degenerate stars.} In: {Plavec} MJ, {Popper} DM,
  {Ulrich} RK (eds) Close Binary Stars: Observations and Interpretation, pp
  251--261

\bibitem[{{Plavec}(1983)}]{1983ApJ...275..251P}
{Plavec} MJ (1983) {Far-ultraviolet emission lines in U Cephei : evidence for a
  hot, turbulent circumstellar envelope.} \apj 275:251--270.
  \doi{10.1086/161530}

\bibitem[{Sana et~al(2012)Sana, {de Mink}, {de Koter}, Langer, Evans, Gieles,
  Gosset, Izzard, Le~Bouquin, and Schneider}]{sana2012}
Sana H, {de Mink} SE, {de Koter} A, et~al (2012) Binary {{Interaction
  Dominates}} the {{Evolution}} of {{Massive Stars}}. Science 337:444.
  \doi{10.1126/science.1223344}

\bibitem[{Sana et~al(2013)Sana, {de Koter}, {de Mink}, Dunstall, Evans,
  {H{\'e}nault-Brunet}, Ma{\'i}z~Apell{\'a}niz, {Ram{\'i}rez-Agudelo}, Taylor,
  Walborn, Clark, Crowther, Herrero, Gieles, Langer, Lennon, and
  Vink}]{sana2013}
Sana H, {de Koter} A, {de Mink} SE, et~al (2013) The {{VLT}}-{{FLAMES Tarantula
  Survey}}. {{VIII}}. {{Multiplicity}} properties of the {{O}}-type star
  population. Astronomy \&amp; Astrophysics, Volume 550, idA107,
  {$<$}NUMPAGES{$>$}22{$<$}/NUMPAGES{$>$} pp 550:A107.
  \doi{10.1051/0004-6361/201219621}

\bibitem[{Scowen et~al(2022)Scowen, Gayley, Ignace, Neiner, Vasudevan,
  Woodruff, Casini, Schultz, {Andersson}, and Wisniewski}]{ScowenTC}
Scowen P, Gayley K, Ignace R, et~al (2022) {The Polstar High Resolution
  Spectropolarimetry MIDEX Mission}. Ap\&SS 367:{\bf Topical Collection}

\bibitem[{{Shrestha} et~al(2018){Shrestha}, {Neilson}, {Hoffman}, and
  {Ignace}}]{2018MNRAS.477.1365S}
{Shrestha} M, {Neilson} HR, {Hoffman} JL, et~al (2018) {Polarization
  simulations of stellar wind bow-shock nebulae - I. The case of electron
  scattering}. \mnras 477(1):1365--1382. \doi{10.1093/mnras/sty724},
  {\href{https://arxiv.org/abs/1712.04958}{{https://arxiv.org/abs/arXiv:1712.04958}}}
  {[astro-ph.SR]}

\bibitem[{{Simmons}(1982)}]{1982MNRAS.200...91S}
{Simmons} JFL (1982) {Analytic treatment of polarization by arbitrary
  scattering mechanisms in circumstellar envelopes. I - Single stars}. \mnras
  200:91--113. \doi{10.1093/mnras/200.1.91}

\bibitem[{{Simmons}(1983)}]{1983MNRAS.205..153S}
{Simmons} JFL (1983) {Analytic treatment of polarization by arbitrary
  scattering mechanisms in circumstellar envelopes. II - Binary stars}. \mnras
  205:153--170. \doi{10.1093/mnras/205.1.153}

\bibitem[{St-Louis et~al(2022)St-Louis, Gayley, Hillier, Ignace, Jones,
  David-Uraz, Richardson, Vink, Peters, Hoffman, Naz\'e, Stevance, Shenar,
  Fullard, Lomax, and Scowen}]{StLouisTC}
St-Louis N, Gayley KG, Hillier DJ, et~al (2022) {UV Spectropolarimetry with
  Polstar: Massive Star Binary Colliding Winds}. Ap\&SS 9999:{\bf Topical
  Collection}

\bibitem[{{Sudar} et~al(2011){Sudar}, {Harmanec}, {Lehmann}, {Yang},
  {Bo{\v{z}}i{\'c}}, and {Ru{\v{z}}djak}}]{2011A&A...528A.146S}
{Sudar} D, {Harmanec} P, {Lehmann} H, et~al (2011) {UX Monocerotis as a W
  Serpentis binary}. \aap 528:A146. \doi{10.1051/0004-6361/201014920},
  {\href{https://arxiv.org/abs/1103.1766}{{https://arxiv.org/abs/arXiv:1103.1766}}}
  {[astro-ph.SR]}

\bibitem[{Tauris et~al(2017)Tauris, Kramer, Freire, Wex, Janka, Langer,
  Podsiadlowski, Bozzo, Chaty, Kruckow, {van den Heuvel}, Antoniadis, Breton,
  and Champion}]{tauris2017}
Tauris TM, Kramer M, Freire PCC, et~al (2017) Formation of {{Double Neutron
  Star Systems}}. The Astrophysical Journal 846:170.
  \doi{10.3847/1538-4357/aa7e89}

\bibitem[{{Taylor} et~al(1991){Taylor}, {Code}, {Nordsieck}, {Anderson},
  {Babler}, {Bjorkman}, {Clayton}, {Magalhaes}, {Meade}, {Schulte-Ladbeck}, and
  {Whitney}}]{1991ApJ...382L..85T}
{Taylor} M, {Code} AD, {Nordsieck} KH, et~al (1991) {First Ultraviolet
  Spectropolarimetry of Hot Supergiants}. \apjl 382:L85. \doi{10.1086/186218}

\bibitem[{{van Rensbergen} et~al(2008){van Rensbergen}, {De Greve}, {De Loore},
  and {Mennekens}}]{2008A&A...487.1129V}
{van Rensbergen} W, {De Greve} JP, {De Loore} C, et~al (2008) {Spin-up and hot
  spots can drive mass out of a binary}. \aap 487(3):1129--1138.
  \doi{10.1051/0004-6361:200809943},
  {\href{https://arxiv.org/abs/0804.1215}{{https://arxiv.org/abs/arXiv:0804.1215}}}
  {[astro-ph]}

\bibitem[{{van Rensbergen} et~al(2010){van Rensbergen}, {De Greve},
  {Mennekens}, {Jansen}, and {De Loore}}]{2010A&A...510A..13V}
{van Rensbergen} W, {De Greve} JP, {Mennekens} N, et~al (2010) {Mass loss out
  of close binaries. Case A Roche lobe overflow}. \aap 510:A13.
  \doi{10.1051/0004-6361/200913272},
  {\href{https://arxiv.org/abs/0908.2021}{{https://arxiv.org/abs/arXiv:0908.2021}}}
  {[astro-ph.SR]}

\bibitem[{{Wang} et~al(2021){Wang}, {Gies}, {Peters}, {G{\"o}tberg},
  {Chojnowski}, {Lester}, and {Howell}}]{2021AJ....161..248W}
{Wang} L, {Gies} DR, {Peters} GJ, et~al (2021) {The Detection and
  Characterization of Be+sdO Binaries from HST/STIS FUV Spectroscopy}. \aj
  161(5):248. \doi{10.3847/1538-3881/abf144},
  {\href{https://arxiv.org/abs/2103.13642}{{https://arxiv.org/abs/arXiv:2103.13642}}}
  {[astro-ph.SR]}

\bibitem[{{Whitney} et~al(2017){Whitney}, {Wood}, {Bjorkman}, {Cohen}, and
  {Wolff}}]{2017ascl.soft11013W}
{Whitney} BA, {Wood} K, {Bjorkman} JE, et~al (2017) {HO-CHUNK: Radiation
  Transfer code}. \eprint{1711.013}

\bibitem[{{Wood} et~al(1996){Wood}, {Bjorkman}, {Whitney}, and
  {Code}}]{1996ApJ...461..828W}
{Wood} K, {Bjorkman} JE, {Whitney} BA, et~al (1996) {The Effect of Multiple
  Scattering on the Polarization from Axisymmetric Circumstellar Envelopes. I.
  Pure Thomson Scattering Envelopes}. \apj 461:828. \doi{10.1086/177105}

\bibitem[{Yoon(2015)}]{yoon2015}
Yoon SC (2015) Evolutionary {{Models}} for {{Type Ib}}/c {{Supernova
  Progenitors}}. Publ Astron Soc Aust 32:e015. \doi{10.1017/pasa.2015.16},
  {\href{https://arxiv.org/abs/1504.01205}{{https://arxiv.org/abs/arXiv:1504.01205}}}

\bibitem[{{Zhao} et~al(2008){Zhao}, {Gies}, {Monnier}, {Thureau}, {Pedretti},
  {Baron}, {Merand}, {ten Brummelaar}, {McAlister}, {Ridgway}, {Turner},
  {Sturmann}, {Sturmann}, {Farrington}, and {Goldfinger}}]{Zhao2008}
{Zhao} M, {Gies} D, {Monnier} JD, et~al (2008) {First Resolved Images of the
  Eclipsing and Interacting Binary {\ensuremath{\beta}} Lyrae}. \apjl
  684(2):L95. \doi{10.1086/592146},
  {\href{https://arxiv.org/abs/0808.0932}{{https://arxiv.org/abs/arXiv:0808.0932}}}
  {[astro-ph]}

\end{thebibliography}


\pagebreak
\section*{Statements \& Declarations}

This research has made use of NASA's Astrophysics Data System and the SIMBAD database, operated at CDS, Strasbourg, France.
The work has also made use of the BeSS database, operated at LESIA, Observatoire de Meudon, France: \url{http://basebe.obspm.fr}.
This research made use of Astropy, \url{http://www.astropy.org} a community-developed core Python package for Astronomy \citep{astropy2013, astropy2018}.

\subsection*{Funding}

GJP gratefully acknowledges support from NASA grant 80NSSC18K0919 and STScI grants HST-GO-15659.002 and HST-GO-15869.001.
RI acknowledges funding support from a grant by the National Science Foundation (NSF), AST-2009412. JLH acknowledges support from NSF under award AST-1816944 and from the University of Denver via a 2021 PROF award.
YN acknowledges support from the Fonds National de la Recherche Scientifique (Belgium), the European Space Agency (ESA) and the Belgian Federal Science Policy Office (BELSPO) in the framework of the PRODEX Programme (contracts linked to XMM-Newton and Gaia).
Scowen acknowledges his financial support by the NASA Goddard Space Flight Center to formulate the mission proposal for \textit{Polstar}.

\subsection*{Competing Interests}
The authors have no relevant financial or non-financial interests to disclose.

\subsection*{Author Contributions} 
All authors shared ideas that motivated this work.  We showed how FUV/NUV spectroscopy and spectropolarimetry is key to understanding the evolution of OB close binaries, especially systemic mass and angular momentum loss and transfer.
All authors contributed to the writing of the paper. KGG computed the evolutionary tracks in the \textit{r-q}~diagram presented in Section~2 and shown in Figure~1. GJP and KGG were the main contributors to Sections~2 and 3. RI and JLH contributed the calculations and text for Section 4. GJP wrote section 5. 

\subsection*{Data Availability}
 Data sharing is not applicable to this article as no datasets were generated or analysed during the current study.

\onecolumn
\centering
\appendix

\section*{Affiliations}

$^{1}${\orgdiv{Department of Physics \& Astronomy,
University of Southern California,
Los Angeles, CA 90089-0484, USA}}

\noindent
$^{2}${\orgdiv{Department of Physics \& Astronomy, University of Iowa, Iowa City, IA, 52242, USA}}

\noindent
$^{3}${\orgdiv{Department of Physics \& Astronomy, East Tennessee State University,
Johnson City, TN 37614, USA}}

$^{4}${\orgdiv{Department of Physics and Astronomy, Western University, London, ON N6A 3K7, Canada}}

$^{5}${\orgdiv{GAPHE, University of Li\`ege, All\'ee du 6 Aout 19c (B5C), 4000-Li\`ege, Belgium}}

$^{6}${\orgdiv{D\'epartement de physique, Universit\'e de Montr\'eal, Campus MIL, 1375 Avenue Th\'er\`ese-Lavoie-Roux
Montr\'eal (Qc)  
H2V 0B3}}

$^{7}${\orgdiv{Department of Physics \& Astronomy, University of Auckland, 38 Princes Street, 1010, Auckland, New Zealand}}

$^{8}${\orgdiv{Armagh Observatory and Planetarium, College Hill, BT61 9DG Armagh, Northern Ireland}}

$^{9}${\orgdiv{Department of Physics and Astronomy, Embry-Riddle Aeronautical University, 3700 Willow Creek Rd, Prescott, AZ, 86301, USA}}

$^{10}${\orgdiv{Department of Physics and Astronomy, 2112 E. Wesley Ave., Denver, CO 80208, USA}}

$^{11}${\orgdiv{Physics Department, United States Naval Academy, 572C Holloway Rd, Annapolis, MD 21402, USA}}

$^{12}${\orgdiv{Anton Pannekoek Institute for Astronomy and Astrophysics, University of Amsterdam, 1090 GE Amsterdam, The Netherlands}}

$^{13}${\orgdiv{Department of Physics \& Astronomy, Michigan State University,567 Wilson Rd., East Lansing, MI 48824, MI}}

$^{14}${\orgdiv{NASA Goddard Space Flight Center, Greenbelt, MD 20771, USA}}

\end{document}